\pdfoutput=1
\documentclass[a4paper,11pt]{article}
\usepackage{amsopn}
\usepackage{amsmath}
\usepackage{amssymb}
\usepackage[pagebackref, colorlinks = true, linkcolor = blue, urlcolor  = 
blue, citecolor = red]{hyperref}
\usepackage{subfig}
\usepackage{bbm}
\usepackage{authblk}
\usepackage{cite}
\usepackage{subfig}
\usepackage[utf8]{inputenc}
\usepackage[OT4]{fontenc}
\usepackage[margin=1in]{geometry}
\usepackage{stmaryrd}

\usepackage{tikz}
\usetikzlibrary{arrows}
\pgfdeclarelayer{back}
\pgfsetlayers{back,main}
\makeatletter
\pgfkeys{%
  /tikz/on layer/.code={
    \def\tikz@path@do@at@end{\endpgfonlayer\endgroup\tikz@path@do@at@end}%
    \pgfonlayer{#1}\begingroup%
  }%
}
\makeatother

\newcommand{\ket}[1]{\ensuremath{|#1\rangle}}
\newcommand{\bra}[1]{\ensuremath{\langle#1|}}
\newcommand{\ketbra}[2]{\ensuremath{\ket{#1}\bra{#2}}}

\newcommand{\Tr}{\mathrm{Tr}}
\newcommand{\tr}{\Tr}
\newcommand{\1}{I}

\newcommand{\ii}{\mathrm{i}}

\newcommand{\E}{\mathbb{E}}
\newcommand{\C}{\mathbb{C}}
\newcommand{\proj}[1]{\ketbra{#1}{#1}}

\newcommand{\diag}{\ensuremath{\mathrm{diag}}}

\newtheorem{theorem}{Theorem}
\newtheorem{definition}{Definition}
\newtheorem{remark}[theorem]{Remark}
\newtheorem{proposition}[theorem]{Proposition}
\newtheorem{lemma}[theorem]{Lemma}
\newtheorem{corollary}[theorem]{Corollary}

\newenvironment{theproof}[1][Proof]{\addvspace{\baselineskip}\noindent\textbf{#1.}
}{\ \hfill\rule{0.5em}{0.5em}\par\addvspace{\baselineskip}}
\usepackage{todonotes}

\title{Almost all quantum channels are equidistant\footnote{In Memory of
Judyta Henek-Pawela: 1986--2016}}

\author[1,2]{Ion Nechita}
\author[3,4]{Zbigniew Pucha{\l}a}
\author[3,5]{{\L}ukasz Pawela\footnote{lpawela@iitis.pl}}
\author[4,6]{Karol {\.Z}yczkowski}
\affil[1]{Zentrum Mathematik, M5, Technische Universit\"at M\"unchen, 
Boltzmannstrasse 3, 85748 Garching, Germany}
\affil[2]{CNRS, Laboratoire de Physique Th\'{e}orique, IRSAMC, Universit\'{e} 
de Toulouse, UPS, F-31062 Toulouse, France}
\affil[3]{Institute of Theoretical and Applied Informatics, Polish Academy of 
Sciences, ulica~Ba{\l}tycka~5, 44-100 Gliwice, Poland}
\affil[4]{Faculty of Physics, Astronomy and Applied Computer Science, 
Jagiellonian University, ulica prof. Stanis\l{}awa \L{}ojasiewicza 11, 
30-348 Krak{\'o}w, Poland}
\affil[5]{Faculty of Applied Physics and Mathematics, National Quantum 
Information Center, Gda{\'n}sk University of Technology, 80-233 Gda{\'n}sk, 
Poland}
\affil[6]{Center for Theoretical Physics, Polish Academy of Sciences, aleja 
Lotników 32/46, 02-668 Warszawa, Poland}

\date{\today}

\begin{document}
\maketitle 

\begin{abstract}
In this work we analyze properties of generic quantum channels in the case of
large system size. We use random matrix theory and free probability to show
that the distance between two independent random channels converges to a
constant value as the dimension of the system grows larger. As a measure of the
distance we use the diamond norm. In the case of a flat Hilbert-Schmidt
distribution on quantum channels, we obtain that the distance converges to
$\frac12 + \frac{2}{\pi}$, giving also an estimate for the maximum success
probability for distinguishing the channels. We also consider the problem of
distinguishing two random unitary rotations.
\end{abstract}

\section{Introduction}

For any linear map $\Phi:M_{d_1}(\mathbb C) \to M_{d_2}(\mathbb C)$, we define
its \emph{Choi-Jamio{\l}kowski matrix} as
\begin{equation}\label{eq:def-J}
J(\Phi) := \sum_{i,j=1}^{d_1} |i\rangle \langle j| \otimes \Phi(|i\rangle \langle j|) \in M_{d_1}(\mathbb C) \otimes M_{d_2}(\mathbb C).
\end{equation}
This isomorphism was first studied by Choi~\cite{cho75a} and
Jamio{\l}kowski~\cite{jamiolkowski1972linear}. Note that some authors prefer to
add a normalization factor of $d_1^{-1}$ if front of the expression for
$J(\Phi)$.  Other authors use the other order for the tensor product factors, a
choice resulting in an awkward order for the space in which $J(\Phi)$ lives.

The rank of the matrix $J(\Phi)$ is called the \emph{Choi rank} of $\Phi$; it is the minimum number $r$ such that the map $\Phi$ can be written as 
$$\Phi(\cdot) = \sum_{i=1}^r A_i \cdot B_i^*,$$
for some operators $A_i,B_i \in M_{d_2 \times d_1}(\mathbb C)$. 

The \emph{diamond norm} was introduced in Quantum Information Theory by Kitaev
\cite[Section 3.3]{kit} as a counterpart to the $1$-norm in the task of
distinguishing quantum channels. First, define the $1 \to 1$ norm of a linear
map $\Phi:M_{d_!}(\mathbb C) \to M_{d_2}(\mathbb C)$ as
$$\|\Phi\|_{1 \to 1} := \sum_{X \neq 0} \frac{\|\Phi(X)\|_1}{\|X\|_1}.$$
Kitaev noticed that the $1 \to 1$ norm is not stable under tensor products (as it can easily be seen by looking at the transposition map), and considered the following ``regularization'':
$$\|\Phi\|_\diamond:= \sup_{n \geq 1}\|\Phi \otimes \operatorname{id}_n\|_{1 \to 1}.$$
In operator theory, the diamond norm was known before as the \emph{completely bounded trace norm}; indeed, the $1 \to 1$ norm of an operator is the $\infty \to \infty$ norm of its dual, hence the diamond norm of $\Phi$ is equal to the completely bounded (operator) norm of $\Phi^*$ (see \cite[Chapter 3]{pau}). 

We shall need two simple properties of the diamond norm. First, note that the
supremum in the definition can be replaced by taking the value $n=d_1$ (recall
that $d_1$ is the dimension of the input Hilbert space of the linear map
$\Phi$); actually, one could also take $n$ equal to the Choi rank of the map
$\Phi$, see \cite[Theorem 3.3]{tim} or \cite[Theorem 3.66]{wat}. Second, using
the fact that the extremal points of the unit ball of the $1$-norm are unit
rank matrices, we always have
$$\|\Phi\|_\diamond = \sup\{\|(\Phi \otimes \operatorname{id}_{d_1})(|x\rangle \langle y |)\|_1 \, : \, x,y \in \mathbb C^{d_1} \otimes \mathbb C^{d_1}, \, \|x\| = \|y\| = 1\}.$$
Moreover, if the map $\Phi$ is Hermiticity-preserving (e.g.~$\Phi$ is the
difference of two quantum channels), one can optimize over $x=y$ in the formula
above, see \cite[Theorem 3.53]{wat}.

Given a map $\Phi$, it is in general difficult to compute its diamond norm.
Computationally, there is a semidefinite program for the diamond norm,
\cite{wat13}, which has a simple form and which has been implemented in various
places (see, e.g.~\cite{qet}). We will bound the diamond norm in terms of the
partial trace of the absolute value of the Choi-Jamio{\l}kowski matrix.

The diamond norm finds applications in the problem of quantum channel
discrimination. Suppose we have an experiment in which our goal is to
distinguish between two quantum channels $\Phi$ and $\Psi$. Each of the
channels may appear with probability $\frac12$. Then, celebrated results by
Helstrom~\cite{helstrom1976quantum}, Holevo~\cite{hol72}, and Kitaev~\cite{kit} give an upper bound on the probability
of correct discrimination
\begin{equation}
p \leq \frac12 + \frac14 \| \Phi - \Psi \|_\diamond.
\end{equation}

The main goal of this work is study the asymptotic behavior of the diamond 
norm of the difference of two independent quantum channels. To achieve this, in Section~\ref{sec:sdp} we find a new upper bound of on 
the diamond norm of a general map. In our case, it has a nice form
\begin{equation}\label{eq:intro-UB}
\| \Phi -\Psi\|_\diamond \leq \| \tr_2 |J(\Phi - \Psi)| \|_\infty.
\end{equation}
Next, in Section~\ref{sec:lower-bound} we prove that the well known lower bound
on the diamond norm, $\|J(\Phi-\Psi)\|_1 \leq \| \Phi-\Psi \|_\diamond$,
converges to a finite value for random independent quantum channels $\Phi$ and
$\Psi$ in the limit $d_{1,2} \to \infty$. We obtain that for channel sampled
from the flat Hilbert-Schmidt distribution, the value of the lower bound is
\begin{equation}
\lim_{d_{1,2}\to\infty} \frac{1}{d_1} \| J(\Phi - \Psi) \|_1 = \frac12 + 
\frac{2}{\pi} \;\; 
\mathrm{a.s.}
\end{equation}
Finally, 
in Section~\ref{sec:upper-bound} we show that the upper bound  \eqref{eq:intro-UB} also converges to 
the same value as the lower bound. From these results, we infer that for independent random quantum channels 
sampled from the Hilbert-Schmidt distribution, we have
\begin{equation}
\lim_{d_{1,2}\to\infty}\|\Phi - \Psi \|_\diamond = \frac12 + \frac{2}{\pi} \;\; 
\mathrm{a.s.}
\end{equation}
In particular, the optimal success probability of distinguishing the two channels (in the asymptotical regime) is 
\begin{equation}
p \leq \frac{1}{2} + \frac{1}{4}\left( \frac12 + \frac{2}{\pi} \right) = \frac{5}{8} + \frac{1}{2\pi} \approx 0.7842.
\end{equation}
Several generalizations of this type of results are gathered in Theorem \ref{thm:main}, the main result of this paper. 

In Sections \ref{sec:depol} and \ref{sec:unitary} we address respectively two similar problems: distinguishing a random quantum channels from the maximally depolarizing channel and distinguishing two random unitary channels. 

\section{Some useful bounds for the diamond norm}\label{sec:sdp}

We discuss in this section some bounds for the diamond norm. For a matrix $X$, we denote by $\sqrt{X^*X}$ and $\sqrt{XX^*}$ its right and 
left absolute values, i.e. 
$$ \sqrt{X^*X} = V \Sigma V^* \qquad \text{ and } \qquad \sqrt{XX^*} = U \Sigma 
U^*,$$ when $X = U \Sigma V^*$ is the SVD of $X$. In the case where $X$ is 
self-adjoint, we obviously have $\sqrt{X^*X} = \sqrt{XX^*}$.

In the result below, the lower bound is well-known, while the upper bound
appeared in a weaker and less general form in \cite[Theorem 2]{jpl}.

\begin{proposition}\label{prop:bound-diamond}
For any linear map $\Phi:M_{d_1}(\mathbb C) \to M_{d_2}(\mathbb C)$, we have
\begin{enumerate}
\item \label{it:prop-diamond-1}
\begin{equation}\label{eq:bound-diamond}
\frac{1}{d_1} \|J(\Phi)\|_1 \leq \|\Phi\|_\diamond \leq \frac{\| 
\operatorname{Tr}_2 \sqrt{J(\Phi)^*J(\Phi)} \|_\infty + \| \operatorname{Tr}_2 
\sqrt{J(\Phi)J(\Phi)^*} \|_\infty}{2}.
\end{equation}

\item \label{it:prop-diamond-2} Above bounds are equal iff the PSD matrices 
$\varphi := \operatorname{Tr}_2 \sqrt{J(\Phi)^*J(\Phi)}$ and $\psi := 
\operatorname{Tr}_2 \sqrt{J(\Phi)J(\Phi)^*}$  are both scalar.
\end{enumerate}
\end{proposition}
\begin{theproof}
We start by proving item~\ref{it:prop-diamond-1}. Consider the semidefinite
programs for the diamond norm given in \cite[Section 3.2]{wat13}:
\begin{center}
  \begin{minipage}[t]{2in}
    \centerline{\underline{Primal problem}}\vspace{-4mm}
    \begin{align*}
    \text{maximize:}\quad &
    \frac{1}{2} \langle X, J(\Phi) \rangle + \frac{1}{2} \langle X^*,J(\Phi)^* \rangle
    \\[2mm]
    \text{subject to:}\quad & 
    \begin{bmatrix}
      \rho_0  \otimes I_{d_2} & X\\
      X^* & \rho_1 \otimes I_{d_2}
    \end{bmatrix}
    \geq 0\\
    & \rho_0,\rho_1\in M_{d_1}^{1,+}(\mathbb C)\\
    & X \in M_{d_1d_2}(\mathbb C)
    \end{align*}
  \end{minipage}
  \qquad
  \begin{minipage}[t]{2in}
    \centerline{\underline{Dual problem}}\vspace{-4mm}
    \begin{align*}
    \text{minimize:}\quad & 
    \frac{1}{2} \| \operatorname{Tr}_2 Y_0 \|_\infty
    + \frac{1}{2} \| \operatorname{Tr}_2 Y_1 \|_\infty\\[2mm]
    \text{subject to:}\quad &
    \begin{bmatrix}
      Y_0 & -J(\Phi)\\
      -J(\Phi)^* & Y_1
    \end{bmatrix}
    \geq 0\\
    & Y_0, Y_1 \in M_{d_1d_2}^+(\mathbb C)
    \end{align*}
  \end{minipage}

  \end{center}
  
  The lower and upper bounds will follow from very simple feasible points for the primal, resp.~the dual problems. Let $J(\Phi) = U \Sigma V^*$ be a SVD of the Choi-Jamio{\l}kowski state of the linear map. For the primal problem, consider the feasible point $\rho_{0,1} = d_1^{-1} I_{d_1}$ and $X = d_1^{-1} UV^*$. The value of the primal problem at this point is 
$$\frac{1}{2d_1} \langle UV^*, UV^* |J(\Phi)| \rangle + \frac{1}{2d_1} \langle VU^*, |J(\Phi) | VU^* \rangle = \frac{1}{d_1} \|J(\Phi) \|_1,$$
showing the lower bound. 

For the upper bound, set $Y_0 = \sqrt{J(\Phi)J(\Phi)^*} = U \Sigma U^*$ and $Y_1
= \sqrt{J(\Phi)^*J(\Phi)} = V \Sigma V^*$, both PSD matrices. The condition in
the dual problem is satisfied:
$$\begin{bmatrix}
      Y_0 & -J(\Phi)\\
      -J(\Phi)^* & Y_1
    \end{bmatrix} 
    = 
    \begin{bmatrix}
      U \Sigma U^* & -U \Sigma V^* \\
      -V \Sigma U^* & V \Sigma V^*
    \end{bmatrix} 
 = 
     \begin{bmatrix}
      U & 0 \\
      0 & V
    \end{bmatrix} \cdot 
      \left( \begin{bmatrix}
      1 & -1 \\
      -1 & 1
    \end{bmatrix} \otimes \Sigma \right) \cdot
       \begin{bmatrix}
      U & 0 \\
      0 & V
    \end{bmatrix} ^*
\geq 0,    $$
and the proof of item~\ref{it:prop-diamond-1} is complete.

To show statement in item~\ref{it:prop-diamond-2} note that the 
lower bound in \eqref{eq:bound-diamond} can be rewritten as
$$\frac{1}{d_1} \|J(\Phi)\|_1 = \frac{1}{d_1} \operatorname{Tr} \varphi = 
\frac{1}{d_1} \operatorname{Tr} \psi,$$
and the two bounds are equal exactly when the spectra of $\varphi$ and $\psi$
are flat. This is also the necessary and sufficient condition for the saturation
of the lower bound, see \cite{kkeg,mkkg}.
\end{theproof}

\begin{corollary}
If the map $\Phi$ is Hermiticity-preserving (i.e.~the matrix $J(\Phi)$ is self-adjoint), the inequality in the statement reads simply
$$\frac{1}{d_1} \|J(\Phi)\|_1 \leq \|\Phi\|_\diamond \leq \| \operatorname{Tr}_2 |J(\Phi)| \|_\infty.$$
\end{corollary}

Let us now characterize the maps $\Phi$ for which the upper bound in \eqref{eq:bound-diamond} is saturated. Since our proof is SDP-based, we use the same technique as in \cite[Theorem 18]{kkeg}.

\begin{proposition}
A map $\Phi$ saturates the upper bound in \eqref{eq:bound-diamond} iff there exist unit vectors $a,b \in \mathbb C^{d_1}$ and a unitary operator $W \in \mathcal U_{d_1d_2}$ with the following properties (we write $J = J(\Phi)$):
\begin{itemize} 
\item The vector $a$ achieves the operator norm for $\operatorname{Tr}_2 \sqrt{JJ^*}$
\item The vector $b$ achieves the operator norm for $\operatorname{Tr}_2 \sqrt{J^*J}$
\item $(aa^* \otimes I_{d_2})W = W(bb^* \otimes I_{d_2})$
\item $J = WP$ for some positive semidefinite operator $P$; in other words, $W$ is the angular part in some polar decomposition of $J$.
\end{itemize}
\end{proposition}
\begin{theproof}
The reasoning follows closely the proof of \cite[Theorem 18]{kkeg}, we only sketch the main lines. Writing the SDP in the standard form (see also \cite[Section 3.2]{wat13} for the notation). Optimal matrices for the primal and the dual program are, respectively
$$A_{opt} = \begin{bmatrix}
\rho_0 & . & . & . \\
. & \rho_1 & . & . \\
. & . & \rho_0 \otimes I_{d_2} & W \\
. & . & W^* & \rho_1 \otimes I_{d_2} 
\end{bmatrix},
B_{opt} = \frac 1 2\begin{bmatrix}
\|\operatorname{Tr}_2 \sqrt{JJ^*} \|_\infty & . & . & . \\
. & \|\operatorname{Tr}_2 \sqrt{J^*J} \|_\infty & . & . \\
. & . & \sqrt{JJ^*} & . \\
. & . & . & \sqrt{J^*J} 
\end{bmatrix},$$
where $.$ denotes an unimportant element. Since strong duality holds for our primal-dual pair \cite[Section 3.2]{wat13}, \emph{complementary slackness} holds and we have
\begin{align*}
\left(\|\operatorname{Tr}_2 \sqrt{JJ^*} \|_\infty I - \operatorname{Tr}_2 \sqrt{JJ^*} \right)\rho_0 &= 0\\
\left(\|\operatorname{Tr}_2 \sqrt{J^*J} \|_\infty I - \operatorname{Tr}_2 \sqrt{J^*J}\right)\rho_1 &= 0\\
U \Sigma U^* R &= U \Sigma V^* (\rho_1 \otimes I_{d_2})\\
V \Sigma V^* R^* &= V \Sigma U^* (\rho_0 \otimes I_{d_2}),
\end{align*}
where $J = U \Sigma V^*$ is the singular value decomposition of $J$. Using an approximation argument, we can assume $J$ (and thus $\Sigma$) is invertible, and thus $W = UV^*$ is unique. We then set $\rho_0 = aa^*$ and $\rho_1 = bb^*$, and the result follows. 
\end{theproof}
\begin{remark}
The upper bound in \eqref{eq:bound-diamond} can be seen as a strengthening of the following inequality $\|\Phi\|_\diamond \leq \|J(\Phi)\|_1$, which already appeared in the literature (e.g.~\cite[Section 3.4]{wat}). Indeed, again in terms of $\varphi$ and $\psi$, we have $\|\varphi\|_\infty \leq \|\varphi\|_1$ and $\|\psi\|_\infty \leq \|\psi\|_1$. The inequality in \eqref{eq:bound-diamond} is much stronger: for example, it is always saturated for tensor product matrices $J = J_1 \otimes J_2$ ($W$ from the result above is also product), whereas the weaker inequality $\|\Phi\|_\diamond \leq \|J(\Phi)\|_1$ is saturated in this case only when $J_1$ has rank one, see \cite{kkeg,mkkg}.
\end{remark}
%
%

\section{Discriminating random quantum channels}

\subsection{Probability distributions on the set of quantum channels}

There are several ways to endow the convex body of quantum channels with probability distributions. In this section, we discuss several possibilities and the relations between them. 

Recall that the Choi-Jamio{\l}kowski isomorphism puts into correspondence a quantum channel $\Phi : M_{d_1}(\mathbb C) \to M_{d_2}(\mathbb C)$ with a bipartite matrix $J(\Phi) \in M_{d_1}(\mathbb C) \otimes M_{d_2}(\mathbb C)$ having the following two properties
\begin{itemize}
\item $J(\Phi)$ is positive semidefinite
\item $\operatorname{Tr}_2 J(\Phi) = I_{d_1}$. 
\end{itemize}
The above two properties correspond, respectively, to the fact that $\Phi$ is
complete positive and trace preserving. Hence, it is natural to consider
probability measures on quantum channels obtained as the image measures of
probabilities on the set of bipartite matrices with the above properties.
Henceforth we will denote the set of all quantum channels as $\Theta(d_1,
d_2)$.

Given some fixed dimensions $d_1,d_2$ and a parameter $s \geq d_1 d_2$, let $G
\in M_{d_1d_2 \times s}(\mathbb C)$ be a random matrix having i.i.d.~standard
complex Gaussian entries; such a matrix is called a \emph{Ginibre random
matrix}. Define then
\begin{align}
\label{eq:Wishart} W &:= GG^* \in M_{d_1}(\mathbb C) \otimes M_{d_2}(\mathbb C) \\
\label{eq:def-partial-normalization}
D &:= \left((\tr_2 W)^{-1/2} \otimes I_{d_2} \right) W \left((\tr_2 W)^{-1/2} \otimes I_{d_2} \right)\in M_{d_1}(\mathbb C) \otimes M_{d_2}(\mathbb C).
\end{align}
The random matrices $W$ and $D$ are called, respectively, \emph{Wishart} and \emph{partially normalized Wishart}. The inverse square root in the definition of $D$ uses the Moore-Penrose convention if $W$ is not invertible; note however that this is almost never the case, since the Wishart matrices with parameter $s$ larger than its size is invertible with unit probability. It is for this reason we do not consider here smaller integer parameters $s$. Note that the matrix $D$ satisfies the two conditions discussed above: it is positive semidefinite and its partial trace over the second tensor factor is the identity:
\begin{align*}
\operatorname{Tr}_2 D &= \operatorname{Tr}_2 \left[ \left((\tr_2 W)^{-1/2} \otimes I_{d_2} \right) W \left((\tr_2 W)^{-1/2} \otimes I_{d_2} \right) \right ] \\
&= (\tr_2 W)^{-1/2}  \left( \operatorname{Tr}_2 W \right) (\tr_2 W)^{-1/2} = I_{d_1}.
\end{align*}
Hence, there exists a quantum channel $\Phi_G$, such that $J(\Phi_G) = D$ (note that $D$, and thus $\Phi$ are functions of the original Ginibre random matrix $G$).

\begin{definition}\label{def:measure-partially-normalized-Wishart}
The image measure of the Gaussian standard measure through the map $G \mapsto \Phi_G$ defined in \eqref{eq:Wishart}, \eqref{eq:def-partial-normalization} and the equation $J(\Phi_G) = D$ is called the \emph{partially normalized Wishart measure} and is denoted by $\gamma^W_{d_1,d_2,s}$.
\end{definition}
Of particular interest is the case $s = d_1d_2$; the measure obtained in this 
case will be called the \emph{Hilbert-Schmidt measure} as it is induced from 
the Hilbert-Schmidt measure on the space of bipartite quantum states by partial 
normalization~\cite{Bruzda09} and will be denoted by $\gamma^{HS}$ (see 
\cite{sommers2004statistical} for the case of random quantum states).

Let us mention here also other measures in the space of quantum operations
discussed in the literature. One can use the Stinespring dilation theorem
\cite{sti}: for any channel $\Phi: M_{d_1}(\mathbb C) \to M_{d_2}(\mathbb C)$,
there exists, for some given $s \leq d_1d_2$, an isometry $V: \mathbb C^{d_1}
\to \mathbb C^{d_2} \otimes \mathbb C^{s}$ such that
\begin{equation}\label{eq:Stinespring}
\Phi(\cdot) = \operatorname{Tr}_2 (V \cdot V^*).
\end{equation}

\begin{definition}\label{def:measure-isometries}
For any integer parameter $s$, let $\gamma^{Haar}_{d_1,d_2,s}$ be the image
measure of the Haar distribution on isometries $V$ through the map in
\eqref{eq:Stinespring}.
\end{definition}

Finally, one can consider the Lebesgue measure on the convex body of quantum
channels, $\gamma^L_{d_1,d_2}$ which leads to the Euclidean geometry of this
set~\cite{Szarek08}. In this work, we shall however be concerned only with the
measure $\gamma^W$ coming from normalized  Wishart matrices. The relations
between all these probability measures on the set of quantum channels shall be
investigated in some future work.

\subsection[The (two-parameter) subtracted Marcenko-Pastur distribution]{The (two--parameter) subtracted Mar\u{c}enko--Pastur distribution}

In this section we introduce and study the basic properties of a two-parameter family of probability measures which will appear later in the paper. This family generalizes the symmetrized Mar\u{c}enko--Pastur distributions from \cite{ppz}, see also \cite{nsp98,dno} for other occurrences of some special cases. Before we start, recall that the Mar\u{c}enko--Pastur (of free Poisson) distribution of parameter $x>0$ has density given by \cite[Proposition 12.11]{nsp}
$$d\mathcal{MP}_x = \max (1-x,0)\delta_0+\frac{\sqrt{4x-(u-1-x)^2}}{2\pi u}1_{[a,b]}(u)\, du,$$
where $a=(\sqrt x - 1)^2$ and $b=(\sqrt x +1)^2$.

\begin{definition}\label{def:SMP-xy}
Let $a,b$ be two free random variables having Mar\u{c}enko--Pastur distributions 
with respective parameters $x$ and $y$. The distribution of the random variable 
$a/x - b/y$ is called the \emph{subtracted Mar\u{c}enko--Pastur distribution} 
with parameters $x,y$ and is denoted by $\mathcal{SMP}_{x,y}$. In other words, 

\begin{equation}\label{eq:def-SMP-xy}
\mathcal{SMP}_{x,y} = D_{1/x} \mathcal{MP}_x \boxplus D_{-1/y}\mathcal{MP}_y,
\end{equation}
where $D_{c} \mathcal{P}$ is a distribution of a random variable $Z' = cZ$ 
provided $Z$ is distributed according to $\mathcal{P}$.
\end{definition}

We have the following result. 

\begin{proposition}\label{prop:SMP-Wishart}
Let $W_x$ (resp.~$W_y$) be two Wishart matrices of parameters $(d,s_x)$ (resp~$(d,s_y)$). Assuming that $s_x/d \to x$ and $s_y/d \to y$ for some constants $x,y >0$, then, almost surely as $d \to \infty$, we have 
$$\lim_{d \to \infty} \|(xd^2)^{-1}W_x - (y d^2)^{-1}W_y\|_1 = \int |u| \, d\mathcal{SMP}_{x,y}(u) =: \Delta(x,y).$$
\end{proposition}
\begin{theproof}
The proof follows from standard arguments in random matrix theory, and from the fact that the Schatten $1$-norm is the sum of the singular values, which are the absolute values of the eigenvalues in the case of self-adjoint matrices.
\end{theproof}

\begin{figure}[ht]
\centering
\subfloat[$x=1$, $y=1$]{\includegraphics{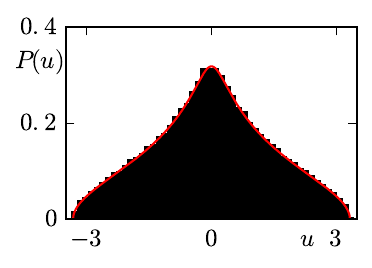}}
\subfloat[$x=1$, $y=2$]{\includegraphics{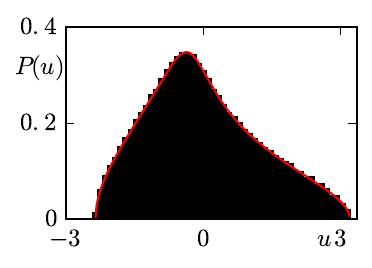}}\\
\subfloat[$x=0.5$, $y=1$]{\includegraphics{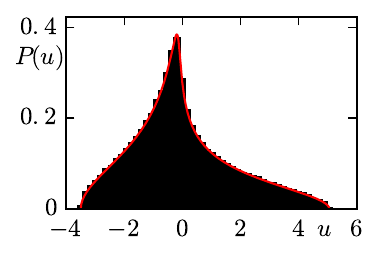}}
\subfloat[$x=0.25$, $y=0.5$]{\includegraphics{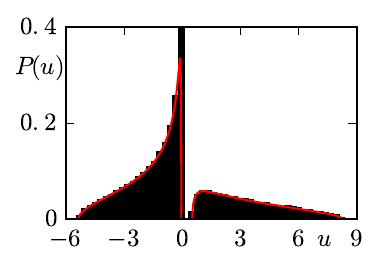}}
\caption{Subtracted Mar\u{c}enko--Pastur distribution for (x, y)=(1,1) (a), (1, 
2) (b), (0.5, 1) (c) and (0.25, 0.5) (d). The red curve is the plot of 
\eqref{eq:smp-density}, while the black histogram corresponds to Monte Carlo 
simulations. Notice the Dirac mass at zero in the last example.}
\label{fig:SMP-xy}
\end{figure}

We gather next some properties of the probability measure $\mathcal{SMP}_{x,y}$. Examples of this distribution are shown in~Fig.~\ref{fig:SMP-xy}.

\begin{proposition}
Let $x,y>0$. Then, 
\begin{enumerate}
\item If $x+y<1$, then the probability measure $\mathcal{SMP}_{x,y}$ has exactly one atom, located at 0, of mass $1-(x+y)$. If $x+y \geq 1$, then $\mathcal{SMP}_{x,y}$ is absolutely continuous with respect to the Lebesgue measure on $\mathbb R$. 
\item Define
\begin{equation}
\begin{split}
a_{x,y} &= (x-y)(2x+y)(x+2y)\\
b_{x,y} &= 2x^3+2y^3 +(x+y)^2 + xy(x+y+2) \\
c_{x,y} &= (x-y)(x+y+1 - 2(x+y)^2)\\
U_{x,y}(u) &= -u^3 a_{x,y}+3 u^2 b_{x,y} + 3 u c_{x,y} + 2(x+y-1)^3\\ %
T_{x,y}(u) &= (x+y-1-u(x-y))^2+3u(y-x+uxy)\\ Y_{x,y}(u) &=
U_{x,y}(u)+\sqrt{\left[ U_{x,y}(u) \right]^2 - 4 \left[ T_{x,y}(u) \right]^3}.
\end{split}
\end{equation}
The support of the absolutely continuous part of $\mathcal{SMP}_{x,y}$ is the set
\begin{equation}\label{eq:smp-support-set}
\{u \, : \, \left[ U_{x,y}(u)\right]^2 - 4 \left[ T_{x,y}(u) \right]^3 \geq
0\}.
\end{equation}
\item On its support, the density of $\mathcal{SMP}_{x,y}$ is given by
\begin{equation}\label{eq:smp-density}
\frac{d\mathcal{SMP}_{x,y}}{du} = \left| \frac{\left[ Y_{x,y}(u) 
\right]^{\frac23}-2^{\frac23} T_{x,y}(u)}{2^{\frac43} \sqrt{3} \pi u \left[ Y_{x,y}(u) \right]^{\frac13}} \right|.
\end{equation}
\end{enumerate}
\end{proposition}
\begin{theproof}
The statement regarding the atoms follows from \cite[Theorem 7.4]{bvo}.  The formula for the density and equation \eqref{eq:smp-support-set} comes from Stieltjes inversion, see e.g.~\cite[Lecture 12]{nsp}. Indeed, since the $R$-transform of the Mar\u{c}enko--Pastur distribution $\mathcal{MP}_x$ reads $R_x(z) = x/(1-z)$, the $R$-transform of the subtracted measure reads
$$R(z) = \frac{1}{1-z/x} - \frac{1}{1+z/y}.$$
The Cauchy transform $G$ of $\mathcal{SMP}_{x,y}$ can be obtained from the 
functional equation
$$
R(G)+1/G=z.
$$
This leads to the following third degree polynomial equation for $G$
$$
z G^3 + [x(1-z)+y(1+z)-1]G^2+(x-y-xyz)G+xy=0.
$$
Using Cardano's formulas for the solutions of a cubic, we can solve this equation and obtain a solution $G(z)$. The 
last step is to perform the Stieltjes inversion
$$
\mathcal{SMP}_{x,y}(u)=-\frac{1}{\pi} \lim_{\varepsilon \to 0} \Im G(u+\ii 
\varepsilon).
$$

\end{theproof}

In the case where $x=y$, some of the formulas from the result above become simpler (see also \cite{ppz}). When $x=y>1/2$, the distribution of $\mathcal{SMP}_{x,x}$ is supported between
$$u_\pm = \pm
\frac{\sqrt{2+10x - x^2 + (x+4)^\frac32 \sqrt{x}}}{\sqrt{2}x} .$$ 
When $x=y\leq 1/2$, $\mathcal{SMP}_{x,x}$ has an atom in $0$ of mass $1-2x$, and its absolutely continuous part is supported on $[u_-,v_-] \cup [v_+,u_+]$, where
$$v_\pm =\pm\frac{\sqrt{2+10x - x^2 - (x+4)^\frac32 \sqrt{x}}}{\sqrt{2}x} .$$ 
Finally, in the case when $x=y=1$, which corresponds to a
flat Hilbert-Schmidt measure on the set of quantum channels, we get that
$\Delta(1,1) = \frac12 + \frac{2}{\pi}$.

\section{The asymptotic diamond norm of the difference of two independent random quantum channels}

We state here the main result of the paper. For the proof, see the following two subsections, each providing one of the bounds needed to conclude. 

\begin{theorem}\label{thm:main}
Let $\Phi$, resp.~$\Psi$, be two \emph{independent} random quantum channels from $\Theta(d_1, d_2)$ having $\gamma^W$distribution with parameters $(d_1,d_2,s_x)$, resp.~$(d_1,d_2,s_y)$. Then, almost surely as $d_{1,2} \to \infty$ in such a way that $s_x/(d_1d_2) \to x$, $s_y/(d_1d_2) \to y$ (for some positive constants $x,y$), and  $d_1 \ll d_2^2$,
$$\lim_{d_{1,2} \to \infty}  \|\Phi - \Psi \|_\diamond = \Delta(x,y) = \int |u| \, d\mathcal{SMP}_{x,y}(u).$$
\end{theorem}

\begin{theproof}
The proof follows from Theorems \ref{thm:lower} and \ref{thm:upper}, which give the same asymptotic value.
\end{theproof}

\begin{remark}
We think that the condition $d_1 \ll d_2^2$ in the statement is purely technical, and could be replaced by a much weaker condition.
\end{remark}

\begin{corollary}
Combining Theorem~\ref{thm:main} with Hellstrom's theorem for quantum 
channels, we get that the optimal probability $p$ of distinguishing two quantum 
channels is equal to:
\begin{equation}
p = \frac58 + \frac{1}{2\pi}.
\end{equation}
Additionally, any maximally entangled state may be used to achieve this value.
\end{corollary}

\subsection{The lower bound}\label{sec:lower-bound}

In this section we compute the asymptotic value of the lower bound in Theorem \ref{thm:main}. Given two random quantum channels $\Phi,\Psi$, we are interested in the asymptotic value of the quantity $d_1^{-1}\|J(\Phi-\Psi)\|_1$. 

\begin{theorem}\label{thm:lower}
Let $\Phi$, resp.~$\Psi$, be two \emph{independent} random quantum channels from
$\Theta(d_1, d_2)$ having $\gamma^W$distribution with parameters
$(d_1,d_2,s_x)$, resp.~$(d_1,d_2,s_y)$. Then, almost surely as $d_{1,2} \to
\infty$ in such a way that $s_x/(d_1d_2) \to x$ and $s_y/(d_1d_2) \to y$ for
some positive constants $x,y$,
$$\lim_{d_{1,2} \to \infty} \frac{1}{d_1} \|J(\Phi-\Psi)\|_1 = \Delta(x,y) = \int |u| \, d\mathcal{SMP}_{x,y}(u).$$
\end{theorem}

The proof of this result (as well as the proof of Theorem \ref{thm:lower}) uses
in a crucial manner the approximation result for partially normalized Wishart
matrices.

\begin{proposition}\label{prop:approximation-partial-normalization}
Let $W\in M_{d_1}(\mathbb C) \otimes M_{d_2}(\mathbb C)$ a random Wishart matrix of parameters $(d_1d_2,s)$, and consider its ``partial normalization'' $D$ as in \eqref{eq:def-partial-normalization}. Then, almost surely as $d_{1,2} \to \infty$ in such a way that $s \sim t d_1d_2$ for a fixed parameter $t>0$, 
$$\big\| D -  (td_1d_2^2)^{-1}W \big\|_{\infty}  = O(d_2^{-2}).$$
\end{proposition}

Note that in the statement above, the matrix $W$ is not normalized; we have
$$\frac{1}{d_1d_2}\sum_{i=1}^{d_1d_2} \delta_{\lambda_i((d_1d_2)^{-1}W)} \to \mathcal{MP}_t,$$
the Mar\u{c}henko--Pastur distribution of parameter $t$. In other words, $W = GG^*$, where $G$ is random matrix of size $d_1d_2 \times s$, having i.i.d.~standard complex Gaussian entries. 

Let us introduce the random matrices 
$$X = (td_1d_2^2)^{-1} \operatorname{Tr}_2 W \quad \text{and} \quad Y =X^{-1/2} \otimes I_{d_2}.$$

The first observation we make is that the random matrix $X$ is also a (rescaled) Wishart matrix. Indeed, the partial trace operation can be seen, via duality, as a matrix product, so we can write
$$X = \frac{1}{td_1d_2^2} \tilde G \tilde G^*,$$
where $\tilde G$ is a complex Gaussian matrix of size $d_1 \times d_1s$; remember that $s$ scales like $td_1d_2$. Since, in our model, both $d_1$, $d_2$ grow to infinity, the behavior of the random matrix $X$ follows from \cite{cny}. 

\begin{lemma}
As $d_{1,2} \to \infty$, the random matrix $\sqrt{t}d_2(X - 	 I_{d_1})$ converges in moments toward a standard semicircular distribution. Moreover, almost surely, the limiting eigenvalues converge to the edges of the support of the limiting distribution:
\begin{align*}
\sqrt{t} d_2 \lambda_{\min}(X- I_{d_1}) &\rightarrow -2\\
\sqrt{t} d_2 \lambda_{\max}(X - I_{d_1}) &\rightarrow 2.
\end{align*}
\end{lemma}
\begin{theproof}
The proof is a direct application of \cite[Corollary 2.5 and Theorem 2.7]{cny}; we just need to check the normalization factors. In the setting of \cite[Section 2]{cny}, the Wishart matrices are not normalized, so the convergence result deals with the random matrices (here $d = d_1$ and $s = td_1d_2^2$)
$$\sqrt{t} d_1 d_2 \left(\frac{\tilde G \tilde G^*}{td_1^2d_2^2} - \frac{I_{d_1}}{d_1} \right) = \sqrt{t}d_2(X - I_{d_1}).$$
\end{theproof}

We look now for a similar result for the matrix $Y$; the result follows by functional calculus. 
\begin{lemma}\label{lem:convergence-Y}
Almost surely as $d_{1,2} \to \infty$, the limiting eigenvalues of the random matrix $\sqrt{t} d_2(Y -  I_{d_1d_2})$ converge respectively to $\pm 1$:
\begin{align*}
\sqrt{t} d_2 \lambda_{\min}(Y - I_{d_1d_2}) &\rightarrow -1\\
\sqrt{t} d_2 \lambda_{\max}(Y - I_{d_1d_2}) &\rightarrow 1.
\end{align*}
\end{lemma}
\begin{theproof}
By functional calculus, we have $\lambda_{\max}(Y) = [\lambda_{\min}(X)]^{-1/2}$, so, using the previous lemma, we get
$$\lambda_{\max}(Y) = \left[ 1-\frac{2}{\sqrt{t} d_2} + o(d_2^{-1}) \right]^{-1/2} = 1 + \frac{1}{2} \frac{2}{\sqrt{t} d_2} + o(d_2^{-1}),$$
and the conclusion follows. The case of $\lambda_{\min}(Y)$ is similar. 
\end{theproof}

We have now all the ingredients to prove Proposition \ref{prop:approximation-partial-normalization}.

\begin{theproof}[Proof of 
Proposition~\ref{prop:approximation-partial-normalization}]
We have
\begin{align*} 
 \big\| D -  (td_1d_2^2)^{-1}W \big\|_{\infty} &= \big\| (td_1d_2^2)^{-1}\left( 
 YWY - W \right)\big\|_{\infty}\\
&= (td_1d_2^2)^{-1} \big\| (Y_i-I)W_iY_i + W_i(Y_i-I)\big\|_{\infty}\\
&\leq (td_1d_2^2)^{-1} \| Y_i-I \|_{\infty} \| W_i \|_{\infty} \left( 
1+\|Y_i\|_{\infty} \right)\\
&= \frac{t^{-3/2}}{d_2^2}\cdot \sqrt{t} d_2 \| Y_i-I \|_{\infty} \cdot  
(d_1d_2)^{-1}\| W_i \|_{\infty} \cdot \left(1+\|Y_i\|_{\infty} \right).
\end{align*}
Note that, almost surely, the three random matrix norms in the last equation above converge respectively to the following finite quantities
\begin{align*}
\sqrt{t} d_2 \| Y_i-I \|_{\infty} &\to 1 \\  
(d_1d_2)^{-1}\| W_i \|_{\infty} &\to (\sqrt{t} +1 )^2 \\
1+ \|Y_i\|_{\infty}  &\to 1.
\end{align*}
The first and the third limit above follow from Lemma \ref{lem:convergence-Y}, while the second one is the Bai-Yin theorem \cite[Theorem 2]{byi} or \cite[Theorem 5.11]{bsi}. 
\end{theproof}


Let us now prove Theorem \ref{thm:lower}.

\begin{theproof}[Proof of Theorem~\ref{thm:lower}]
The result follows easily by approximating the partially normalized Wishart matrices with scalar normalizations. By the triangle inequality, with $D_x:= J(\Phi)$ and $D_y := J(\Psi)$, we have
\begin{align*}
&\left| \frac{1}{d_1} \|D_x- D_y\|_1 -  \frac{1}{d_1} \|(xd_1d_2^2)^{-1}W_x- (yd_1d_2^2)^{-1}W_y\|_1 \right |  \\
& \qquad \qquad \leq \frac{1}{d_1} \|D_x- (xd_1d_2^2)^{-1}W_x\|_1 + \frac{1}{d_1} \|D_y- (yd_1d_2^2)^{-1}W_y\|_1 \\
& \qquad \qquad \leq d_2 \|D_x- (xd_1d_2^2)^{-1}W_x\|_\infty + d_2 \|D_y- (yd_1d_2^2)^{-1}W_y\|_\infty.
 \end{align*}
The conclusion follows from Propositions \ref{prop:SMP-Wishart} and \ref{prop:approximation-partial-normalization}.
\end{theproof}

\subsection{The upper bound}\label{sec:upper-bound}
The core technical result of this work consists of deriving the asymptotic value of the 
upper bound in Theorem \ref{thm:main}. Given two random quantum channels 
$\Phi,\Psi$, we are interested in the asymptotic value of the quantity 
$\|\operatorname{Tr}_2 | J(\Phi-\Psi)|\|_\infty$. 

\begin{theorem}\label{thm:upper}
Let $\Phi$, resp.~$\Psi$, be two \emph{independent} random quantum channels from $\Theta(d_1, d_2)$ having $\gamma^W$distribution with parameters $(d_1,d_2,s_x)$, resp.~$(d_1,d_2,s_y)$. Then, almost surely as $d_{1,2} \to \infty$ in such a way that $s_x/(d_1d_2) \to x$, $s_y/(d_1d_2) \to y$ (for some positive constants $x,y$), and  $d_1 / d_2^2 \to 0$,
$$\lim_{d_{1,2} \to \infty}  \|\operatorname{Tr}_2|J(\Phi-\Psi)|\|_\infty = \Delta(x,y) = \int |u| \, d\mathcal{SMP}_{x,y}(u).$$
\end{theorem}

The proof of Theorem~\ref{thm:upper} is presented at the end of this Section.
It is based on the following lemma which appears in \cite{dav88}; see also
\cite[Eq.~(5.10)]{bha94} or \cite[Chapter X]{bha}.

\begin{lemma}\label{lem:absolute-value-bound}
For any matrices $A,B$ of size $d$, the following holds:
\begin{equation}
\|\ |A|-|B|\ \|_{\infty} \leq C \log d \ \|A - B\|_{\infty},
\end{equation}
for a universal constant $C$ which does not depend on the dimension $d$.
\end{lemma}
For the sake of completeness, we give here a proof, relying on a similar 
estimate for the Schatten classes proved in \cite{dav88}.

\begin{theproof}
Using \cite[Theorem 8]{dav88}, we have, for any $p \in [2,\infty)$:
\begin{align*}
\|\ |A|-|B|\ \|_\infty &\leq \|\ |A|-|B|\ \|_p \\
&\leq 4(1+cp) \| A-B \|_p \\
&\leq 4(1+cp)d^{1/p} \| A-B \|_\infty,
\end{align*}
for some universal constant $c \geq 1$. Choosing $p=\log d$ gives the desired bound, for $d$ large enough. The case of small values of $d$ is obtained by a standard embedding argument. 
\end{theproof}

\begin{theproof}[Proof of Theorem \ref{thm:upper}]
Using the triangle inequality and Lemma \ref{lem:absolute-value-bound}, we first prove an approximation result (as before, we write $D_x := J(\Phi)$ and $D_y :=J(\Psi)$):
\begin{align*}
\big| \ \|\ \tr_2|D_x&-D_y|\ \|_\infty -  \|\ \tr_2|(xd_1d_2^2)^{-1}W_x-(yd_1d_2^2)^{-1}W_y| \ \|_\infty \ \big| \\
&\qquad \leq \big\|   \tr_2|D_x-D_y| - 
\tr_2|(xd_1d_2^2)^{-1}W_x-(yd_1d_2^2)^{-1}W_y| \big\|_{\infty} \\
&\qquad = \big\|   \tr_2\left(|D_x-D_y| -  
|(xd_1d_2^2)^{-1}W_x-(yd_1d_2^2)^{-1}W_y| \right) \big\|_{\infty} \\
&\qquad \leq d_2 \big\| \  |D_x-D_y| -  
|(xd_1d_2^2)^{-1}W_x-(yd_1d_2^2)^{-1}W_y| \  \big\|_{\infty} \\
&\qquad \leq C d_2 \log(d_1d_2) \big\|  (D_x-D_y) -  
((xd_1d_2^2)^{-1}W_x-(yd_1d_2^2)^{-1}W_y)  \big\|_{\infty} \\
&\qquad \leq C d_2 \log(d_1d_2) \left(\big\|   D_x - (xd_1d_2^2)^{-1}W_x \big\|_\infty + \big\| D_y - (yd_1d_2^2)^{-1}W_y  \big\|_\infty \right)\\
&\qquad =  \frac{\log(d_1d_2)}{d_2} O(1) \to 0,
\end{align*}
where we have used Proposition \ref{prop:approximation-partial-normalization} and the fact that $d_1 \ll d_2^2 \implies \log(d_1) \ll d_2$. This proves the approximation result, and we focus now on the simpler case of Wishart matrices. 
Let us define
\begin{align*}
Z&:=(xd_1d_2)^{-1}W_x-(yd_1d_2)^{-1}W_y \\
\tilde Z_1&:= \operatorname{tr}_2(|Z|) = \operatorname{Tr}_2|(xd_1d_2^2)^{-1}W_x-(yd_1d_2^2)^{-1}W_y|
\end{align*}
It follows from \cite[Proposition 4.4.9]{hpe} that the random matrix $Z$ converges almost surely (see Appendix \ref{app:partial-trace-unitarily-invariant} for the definition of almost sure convergence for a sequence of random matrices) to a non-commutative random variable having distribution $\mathcal{SMP}_{x,y}$, see \eqref{eq:def-SMP-xy}. Moreover, using a standard strong convergence argument \cite{mal}, the extremal eigenvalues of $Z$ converge almost surely to the extremal points of the support of the limiting probability measure $\mathcal{SMP}_{x,y}$. Hence, the almost sure convergence extends from the traces of the powers of $Z$ to any continuous bounded function (on the support of $\mathcal{SMP}_{x,y}$), in particular to the absolute value, i.e.~to $|Z|$. 
From Proposition \ref{prop:partial-trace-flat}, the asymptotic spectrum of the random matrix $\tilde Z_1$ is flat, with all the eigenvalues being equal to 
$$ a = \lim_{d_1,d_2 \to \infty} \mathbb E \frac{\operatorname{Tr} |(xd_1d_2)^{-1}W_x-(yd_1d_2)^{-1}W_y|}{d_1d_2} = \int |u| d\mathcal{SMP}_{x,y}(u),$$
which, by Proposition \ref{prop:SMP-Wishart}, is equal to $\Delta(x,y)$, finishing the proof. 

\end{theproof}

\section{Distance to the depolarizing channel}\label{sec:depol}

In this section we derive the asymptotic distance between a random quantum
channel $\Phi$ and the \emph{maximally depolarizing channel}
$$\Psi_\text{dep}: M_{d_1}(\mathbb C) \to M_{d_2}(\mathbb C), \qquad \Psi_\text{dep}(X) = \frac{\operatorname{Tr}(X)}{d_2} I_{d_2}.$$

Let us define the function $g:(1/4, \infty) \to (0,\infty)$
\begin{align*}
g(x)&:= \frac{3}{2}-x +\frac{\sqrt{4 x-1} (2 x+1)}{2 \pi  x} \\
&\qquad - \frac{1}{\pi} \left( (x-1) \operatorname{arctan}\left(\frac{3 x-1}{(x-1) \sqrt{4 x-1}}\right)+ \operatorname{arctan}\left(\sqrt{4 x-1}\right)+ x \operatorname{arctan}\left(\frac{1}{\sqrt{4 x-1}}\right)\right).
\end{align*}

\begin{theorem}
Let $\Phi$ a random quantum channel from $\Theta(d_1,d_2)$ having distribution $\gamma^W$ with parameters $(d_1,d_2,s_x)$. Then, almost surely as $d_1, d_2 \to \infty$ and $s_x \sim xd_1d_2$, we have
\begin{equation}\label{eq:distance-depol}
\lim_{d_1,d_2 \to \infty} \|\Phi - \Psi_\text{dep}\|_\diamond = \int \left| 
\frac{u}{x} - 1 \right| d\mathcal{MP}_x(u) = \begin{cases}
2(1-x) &\qquad \text{ if } x \in (0,1/4]\\
g(x) &\qquad \text{ if } x \in (1/4,1)\\
g(x)+x-1 &\qquad \text{ if } x \in [1,\infty).
\end{cases}
\end{equation}
In the case $x=1$, the limit above reads $3\sqrt 3/(2\pi)$.
\end{theorem}
\begin{remark}
We plot in Figure \ref{fig:distance-depol} the value of the limit in \eqref{eq:distance-depol} as a function of $x$. One can show that the limit is a decreasing function of $x$, converging to $0$ as $x \to \infty$. The function behaves as $8/(3\pi)x^{-1/2}$ as $x \to \infty$. 
\begin{figure}[ht]
\centering\includegraphics[scale=0.8]{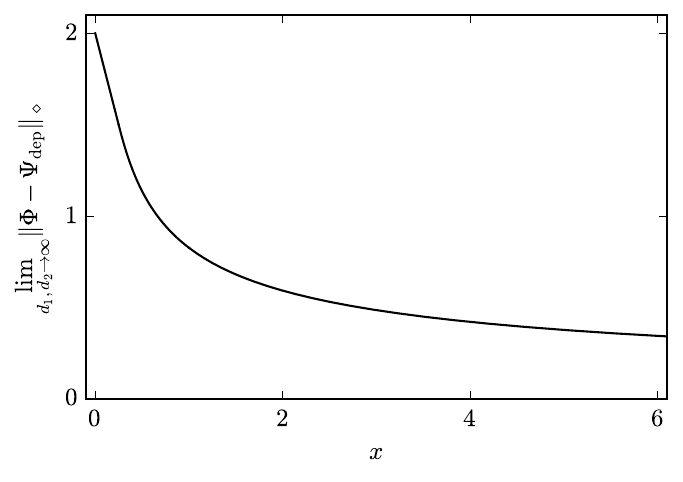}
\caption{The asymptotic diamond-norm distance between a random quantum channel and the maximally depolarizing channel as a function of the channel parameter $x$.}\label{fig:distance-depol}
\end{figure}
\end{remark}
\begin{theproof}
We analyze separately the lower bound and the upper bound from Proposition \ref{prop:bound-diamond}. First, let us denote by $D_x$ the Choi-Jamio{\l}kowski matrix of the channel $\Phi$, and note that $J(\Psi_\text{dep}) = d_2^{-1}I_{d_1d_2}$. For the lower bound, first show that we can approximate the random matrix $D_x$ by a rescaled Wishart matrix:
\begin{align*}\left| \frac{1}{d_1} \|D_x - d_2^{-1}I_{d_1d_2}\|_1 - \frac{1}{d_1} \|(xd_1d_2^2)^{-1}W_x - d_2^{-1}I_{d_1d_2}\|_1 \right| &\leq \frac{1}{d_1} \|D_x-(xd_1d_2^2)^{-1}W_x \|_1 \\
&\leq d_2 \|D_x-(xd_1d_2^2)^{-1}W_x \|_\infty,
\end{align*}
which converges almost surely to 0, by Proposition \ref{prop:approximation-partial-normalization}. The quantity with which we approximate is then
\begin{equation}\label{eq:distance-depol-LB}
\frac{1}{d_1} \|(xd_1d_2^2)^{-1}W_x - d_2^{-1}I_{d_1d_2}\|_1 = \frac{1}{d_1d_2} \sum_{i=1}^{d_1d_2} |\lambda_i[(xd_1d_2)^{-1}W_x - I_{d_1d_2}]|.
\end{equation}
The quantity above converges almost surely, as $d_1d_2 \to \infty$, towards
$$\int \left| \frac{u}{x} - 1 \right| d\mathcal{MP}_x(u).$$

Let us now show that the upper bound from Proposition \ref{prop:bound-diamond} converges to the same quantity. We follow the same steps as in the proof of Theorem \ref{thm:upper}: we first approximate the matrix $D_x$ by a rescaled Wishart random matrix, and then we argue that the partial trace appearing in the bound has ``flat'' eigenvalues, allowing us to replace the operator norm by the normalized trace. For the approximation step, we get, using again Proposition \ref{prop:approximation-partial-normalization},
\begin{align*}
\big| \ \|\ \tr_2|D_x&-d_2^{-1}I_{d_1d_2}|\ \|_\infty -  \|\ \tr_2|(xd_1d_2^2)^{-1}W_x-d_2^{-1}I_{d_1d_2}| \ \|_\infty \ \big| \\
&\qquad \leq \big\|   \tr_2\left(|D_x-d_2^{-1}I_{d_1d_2}| -  |(xd_1d_2^2)^{-1}W_x-d_2^{-1}I_{d_1d_2}| \right) \big\|_\infty \\
&\qquad \leq d_2 \big\| \  |D_x-d_2^{-1}I_{d_1d_2}| -  |(xd_1d_2^2)^{-1}W_x-d_2^{-1}I_{d_1d_2}| \  \big\|_\infty \\
&\qquad \leq C d_2 \log(d_1d_2)\big\|   D_x - (xd_1d_2^2)^{-1}W_x \big\|_\infty \\
&\qquad =  \frac{\log(d_1d_2)}{d_2} O(1) \to 0 \quad \text{almost surely.}
\end{align*}
We focus now on the quantity $\|\ \tr_2|(xd_1d_2^2)^{-1}W_x-d_2^{-1}I_{d_1d_2}| \ \|_\infty$. From Proposition \ref{prop:partial-trace-flat}, the spectrum of the random matrix $\tr_2|(xd_1d_2^2)^{-1}W_x-d_2^{-1}I_{d_1d_2}|$ is flat, so its operator norm has the same limit as $d_1^{-1} \operatorname{Tr} |(xd_1d_2^2)^{-1}W_x-d_2^{-1}I_{d_1d_2}|$,
which is the same as \eqref{eq:distance-depol-LB}, finishing the proof.
\end{theproof}

\section{Distance to the nearest unitary channel}\label{sec:nearest-unitary}
In this section we consider an asymptotic distance between a random quantum 
channel  $\Phi:M_{d }(\mathbb C) \to M_{d}(\mathbb C)$ and a unitary channel. 
First we note, that if a quantum channel $\Phi$ is an interior point of the set 
of channels then, the best distinguishable one $\Psi$ is some unitary 
channel~\cite{puchala2015exploring}. Below we show, that in the case of $d \to 
\infty$ almost all quantum channels are perfectly distinguishable from any 
unitary channel. To see it we write 
\begin{equation}
\begin{split}
\min_{\Psi_U}  \| \Phi - \Psi_U \|_\diamond 
&\geq \frac{1}{d} \min_{\Psi_U}  \| J(\Phi) - J(\Psi_U) \|_1
= \min_{U}  \frac{1}{d} \| J(\Phi) - \ketbra{U}{U}  \|_1
\geq  \min_{ \ket{x} } \frac{1}{d} \| J(\Phi) -  d \ketbra{x}{x}  \|_1 \\
&\geq  \min_{ \ket{x} }  2(1  -  F(J(\Phi)/d,\ketbra{x}{x})  ) 
= 2 - 2 \|J(\Phi)/d \|_{\infty}. 
\end{split}
\end{equation}
In the above we have used the inequality between diamond norm and the trace 
norm of Choi-Ja\-mioł\-kow\-ski matrices, see 
Proposition~\ref{prop:bound-diamond}, and next the Fuchs - van de Graaf 
inequality~\cite{fuchs1999cryptographic} 
involving trace norm and fidelity function $F(\rho,\sigma) = (\tr 
\sqrt{\sqrt{\rho} \sigma \sqrt{\rho}} )^2$.
Next we use the fact, that the largest eigenvalue os matrix $J(\Phi)/d$ 
tends to 0 almost surely. 

\section{Distance between random unitary channels}\label{sec:unitary}

We consider in this section the problem of distinguishing two unitary channels, 
\begin{equation}\label{eq:unitary-channels}
\Phi(X) = UXU^* \qquad \text{and} \qquad \Psi(X) = VXV^*,
\end{equation}
where $U,V$ are two $d \times d$ unitary operators. The diamond norm of the 
difference $\|\Phi - \Psi\|_\diamond$ has already been considered in the 
literature, and we gather below the results from~\cite[Theorem 3.57]{wat} and 
\cite[Theorem 12]{jkp}.

\begin{proposition}
For any two unitary operators $U,V$, the diamond norm of the difference of the unitary channels induced by $U,V$ is given by
\begin{itemize}
\item $\|\Phi - \Psi\|_\diamond = 2\sqrt{1-\nu(U^*V)^2}$, where $\nu(U^*V)$ is the smallest absolute value of an element in the numerical range of the unitary operator $U^*V$. In other words, $\nu(U^*V)$ is the radius of the largest open disc centered at the origin which does not intersect the convex hull of the eigenvalues of $U^*V$ (i.e.~the numerical range). 
\item $\|\Phi - \Psi\|_\diamond = 2R(U^*V)$, where $R(U^*V)$ is the radius of the smallest disc (not necessarily centered at the origin) containing all the eigenvalues of $U^*V$. 
\item Let $2\alpha$ be the smallest arc containing the spectrum of $U^*V$. Then, 
$$\|\Phi - \Psi\|_\diamond = \begin{cases}
2 \sin \alpha, \qquad &\text{ if } \alpha < \pi/2\\
2, \qquad &\text{ if } \alpha \geq \pi/2.
\end{cases}$$
\end{itemize}
\end{proposition}

\begin{figure}[ht]
\centering\includegraphics[scale=0.7]{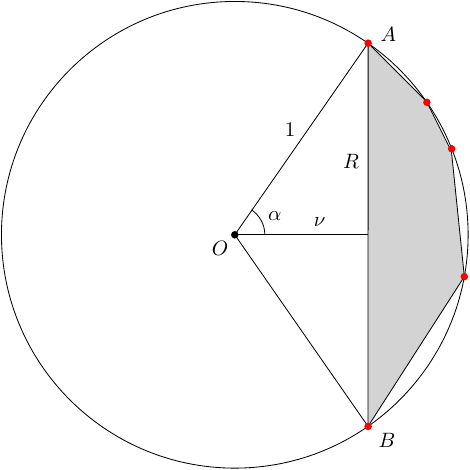} \qquad\qquad\qquad\qquad
\centering\includegraphics[scale=0.7]{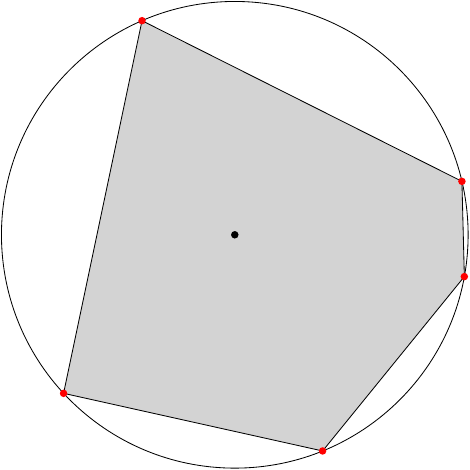}

\caption{The eigenvalues of the unitary operator $A=U^*V$ (red dots, here
$d=5$) and its numerical range $\mathcal{N}(A)$(gray filled area). On the left,
the eigenvalues span an arc of length smaller than $\pi$, so the quantities
$\nu$ and $R$ are non-trivial. On the right, the eigenvalues span an arc larger
than half a circle, so the origin belongs to the numerical range; here, $\nu =
0$ and $R=1$.}\label{fig:nu-R}
\end{figure}

We represent in Figure \ref{fig:nu-R} the eigenvalues of the operator $W := U^*V$ and the numerical range of $W$. Recall that the \emph{numerical range} of an operator $A$ is the set 
$$\mathcal N(A) = \{\langle x, Ax \rangle \, : \, x \in \mathbb C^d, \, \|x\| = 1\}.$$
The numerical range is a convex body \cite[Chapter 1]{hjo91}, and in the case
where $A$ is a normal operator ($AA^* =A^*A$) it coincides with the convex hull
of the spectrum. One remarkable fact about the results in the proposition above
is that two unitary operations $\Phi$ and $\Psi$ become \emph{perfectly}
distinguishable as soon as the convex hull of the eigenvalues of $U^*V$
contains the origin~\cite[Theorem 3.57]{wat}, \cite[Theorem 12]{jkp}.

We consider next random unitary operators $U,V$. We analyze Haar-distributed operators and then the case where $U$ and $V$ are sampled from the distribution of two independent unitary Brownian motions stopped at different times. For independent, Haar-distributed unitary operators, in the limit of large dimension, the corresponding channels become perfectly distinguishable. 

\begin{proposition}
Let $U,V \in \mathcal U(d)$ be two independent random variables, at least one of them being Haar-distributed. Then, with overwhelming probability as $d \to \infty$, the quantum channels $\Phi$ and $\Psi$ from \eqref{eq:unitary-channels} become perfectly distinguishable: for $d$ large enough,
$$\mathbb P\left[ \|\Phi - \Psi\|_\diamond = 2 \right] \geq 1-\exp\left(-\frac{\log 2}{2}d^2 \right).$$
\end{proposition}

\begin{remark}
The statement above includes the case where $U$ is a Haar-distributed random unitary matrix, and $V$ is the identity operator (hence, $\Psi$ is the identity channel).
\end{remark}

\begin{theproof}
From the hypothesis and the left / right invariance of the Haar distribution, it follows that the random matrix $W = U^*V$ is Haar-distributed. The estimate follows from \cite[Section 3.1]{bab}, where the probability of a Haar unitary matrix not having any eigenvalues in a given arc is related to a Toeplitz determinant, see equation (3.1) in  \cite{bab}. 
\end{theproof}

Let us now consider the case where the operators $U$ and $V$ are elements of two independent unitary Brownian motion processes. We shall not give the definition of this process, referring the reader to e.g.~\cite{bia95,bia97,rai97}. We shall only need here the following result of Biane, giving the asymptotic support of a unitary Brownian motion stopped at time $t$. 

\begin{proposition}\cite[Proposition 10]{bia97}
Let $(U_t)_{t \geq 0}$ be a unitary Brownian motion on $\mathcal U(d)$ starting at the identity. Then, asymtptically as $d \to \infty$, the support of the eigenvalue distribution (on the unit circle) of the operator $U_t$ is the full circle if $t \geq 4$ and the arc
$$\left\{\exp(i\alpha) \, : \, |\alpha| \leq \frac{1}{2}\sqrt{t(4-t)} + \arccos(1-t/2) \right\} \qquad \text{ if }0 \leq t <4.$$ 
\end{proposition}

As a direct application of this result, we obtain the diamond norm of the difference of two unitary quantum channels stemming from independent unitary Brownian motions.

\begin{figure}[ht]
\centering\includegraphics[scale=0.7]{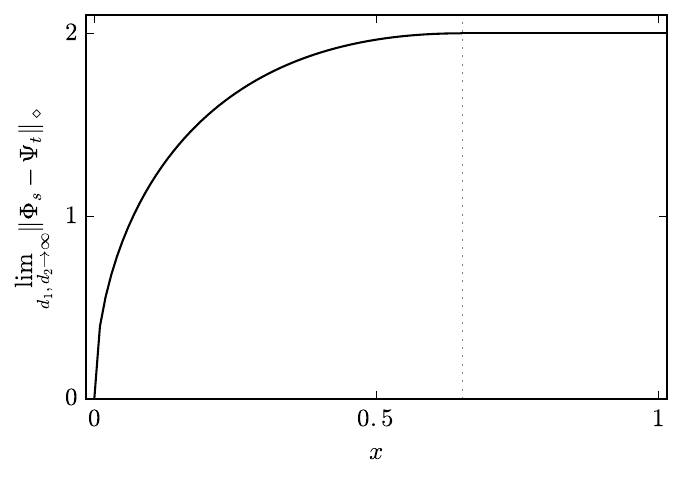} 
\caption{The diamond norm of a difference of two random unitary channels coming from two independent unitary Brownian motions stopped at times $s$ and $t$, as a function of $s+t$.}
\label{fig:uBm}
\end{figure}

\begin{proposition}
Let $(U_s)_{s \geq 0}$ and $(V_t)_{t \geq 0}$ two independent unitary Brownian motions and consider the random unitary quantum channels $\Phi_s$ and $\Psi_t$ from \eqref{eq:unitary-channels} obtained from the operators $U_s$ and respectively $V_t$. Then, almost surely,
$$\lim_{d \to \infty} \|\Phi_s - \Psi_t\|_\diamond = \begin{cases}
2 \sin \left[ \frac{1}{2}\sqrt{(s+t)(4-s-t)} + \arccos(1-(s+t)/2)\right] , \quad &\text{ if } s+t < \tau\\
2, \quad &\text{ if } s+t \geq \tau,
\end{cases}$$
where $\tau \approx 0.6528$ is the unique solution of the equation 
$$\frac{1}{2}\sqrt{t(4-t)} + \arccos(1-t/2) = \pi/2$$
on $(0, 4)$. 
\end{proposition}
\begin{theproof}
The proof is an easy consequence of Biane's result (more precisely, of its ``strong'' formulation from \cite[Theorem 1.1]{cdk}), once we notice that the random unitary matrix $U_s^*V_t$ has the same distribution as $W_{s+t}$, where $W_\cdot$ is another unitary Brownian motion. We plot the diamond norm as a function of $s+t$ in Figure \ref{fig:uBm}. 
\end{theproof}

\section{Concluding remarks} 
\begin{figure}[!h]
\centering\def \size {3}
\def \paddedsize {3.2}

\def \rhox {0}
\def \rhoy {3}
\def \rhodownx {2.8284}
\def \rhodowny {-1}

\def \sigmax {-1.5}
\def \sigmay {-0.33}

\tikzstyle{reverseclip}=[insert path={(current page.north east) --
  (current page.south east) --
  (current page.south west) --
  (current page.north west) --
  (current page.north east)}
]

\begin{tikzpicture}
\draw[line width=2] (-\size,\paddedsize) -- (\size,\paddedsize);
\draw[line width=2] (-\size,-\paddedsize) -- (\size,-\paddedsize);
\draw[line width=2] (-\size,-\paddedsize) arc (270:90:\size cm and \paddedsize 
cm);
\draw[line width=2] (\size,-\paddedsize) arc (-90:90:\size cm and \paddedsize 
cm);

\draw (-\size,0) arc (180:360:\size cm and 0.5cm);
\draw (-\size,0) arc (180:0:\size cm and 0.5cm);

\draw (0,0) circle (\size cm);

\node[draw=none] at (-3.7,-2.3) {\LARGE$\Theta(d, d)$};

\node
[label={[xshift=0.4cm, 
yshift=-0.65cm]\large$\Psi_U$},draw,fill=black,circle,inner
 sep=1pt,minimum size=5pt] at (-6,0) {};

\node [label={[xshift=-0.3cm, 
yshift=-0.8cm]\large$\Psi$},draw,fill=black,circle,inner sep=0pt,minimum
size=5pt] at (\rhox, \rhoy) {};

\node [label={[xshift=0.4cm, 
yshift=-0.42cm]\large$\Phi$},draw,fill=black,circle,inner 
sep=0pt,minimum size=5pt] at (\rhodownx, \rhodowny) {};

\node [label=left:\large$\Phi_\mathrm{dep}$,draw,fill=black,circle,inner 
sep=0pt,minimum size=5pt] at (0,0) {};

\draw [dashed](\rhox,\rhoy) -- (\rhodownx, \rhodowny) 
node[midway,above,font=\scriptsize,rotate=105] {\large$\Delta$};

\draw [dashed](\rhox,\rhoy) -- (-6, 0) 
node[midway,above,font=\scriptsize,rotate=25] {\large$a$};

\draw [dashed](\rhox,\rhoy) -- (0, 0) 
node[midway,above,font=\scriptsize,rotate=90] {\large$r$};

\draw [dashed](\rhodownx,\rhodowny) -- (0, 0) 
node[near start,below,font=\scriptsize,rotate=-15] {\large$r$};

\end{tikzpicture} \caption{Sketch of the set $\Theta(d, d)$ of
all channels acting on $d$-dimensional states. A generic channel $\Phi$ belongs
to a sphere of radius $r=3\sqrt{3}/2\pi$, centered at the maximally
depolarizing channel, $\Phi_\mathrm{dep}$ in the metric induced by the diamond
norm. The distance between generic channels, $\Phi, \Psi$ is $\Delta=1/2 +
2/\pi$, while the distance to the nearest unitary channel reads $a=2$.}
\label{fig:set-of-channels}
\end{figure}
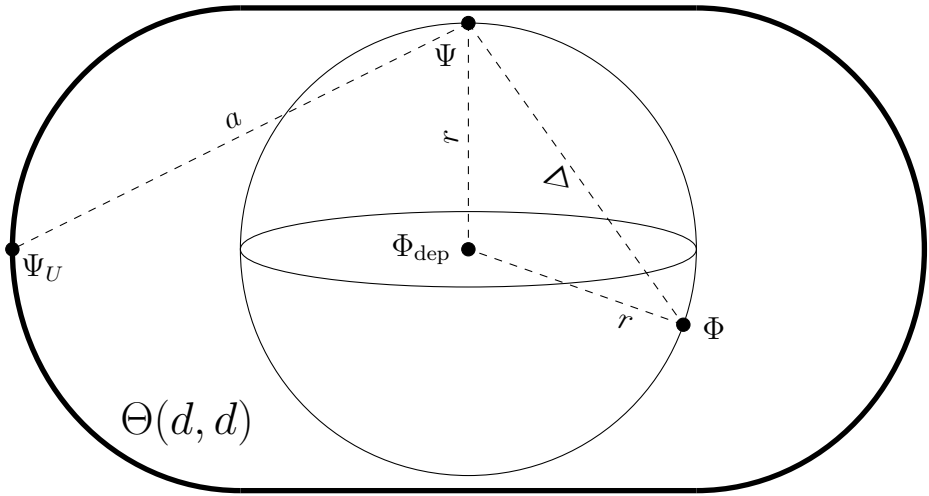

\begin{figure}[!h]
\centering\includegraphics{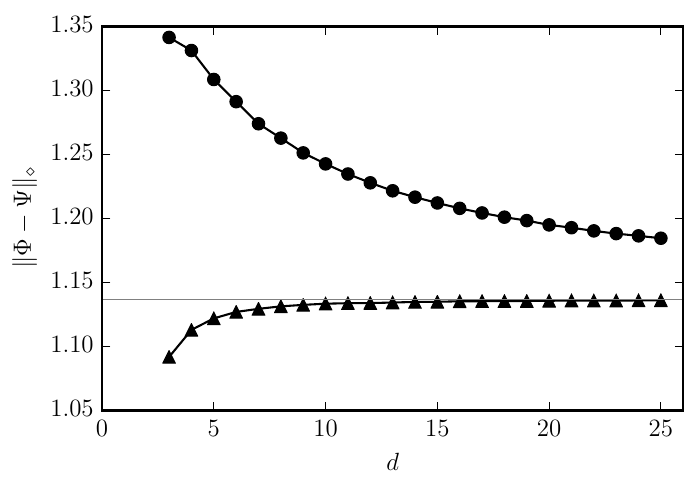} \caption{The convergence of upper
(circles) and lower (triangles) bounds on the distance between two random
quantum channels sampled from the Hilbert-Schmidt distribution ($d_1=d_2=d$).
The results were obtained via Monte Carlo simulation with 100 samples for each
data point.}\label{fig:convergence}
\end{figure}

In this work we analyzed properties of generic
quantum channels concentrating on the case of large system size. Using tools
provided by the theory of random matrices and the free probability calculus we
showed that the diamond norm of the difference between two random channels
asymptotically tends to a constant specified in Theorem \ref{thm:main}. In the
case of channels corresponding to the simplest case $x=y=1$, the limit value of
the diamond norm of the difference is $\Delta(1,1) = 1/2 + 2/\pi$. Based on
these results, in Fig.~\ref{fig:set-of-channels} we provide a sketch of the set
of quantum channels. In Fig.~\ref{fig:convergence} we illustrate the
convergence of the upper and lower bound to the value $1/2 + 2/\pi$. This
statement allows us to quantify the mean distinguishability between two random
channels

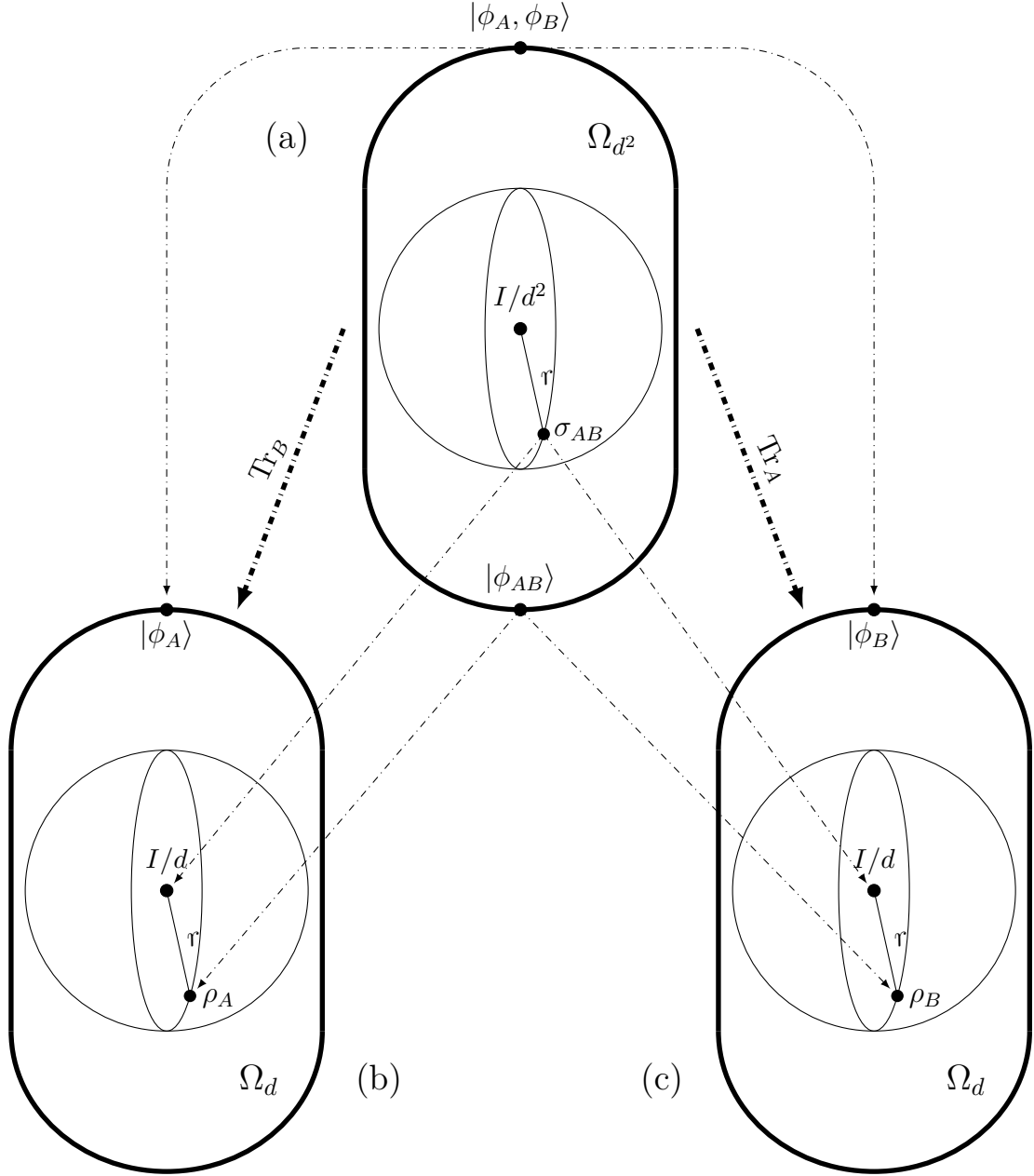
\begin{figure}[!h]
\centering\def \size {2}
\def \paddedsize {2.2}

\def \sigmax {-1.5}
\def \sigmay {-0.33}

\tikzstyle{reverseclip}=[insert path={(current page.north east) --
  (current page.south east) --
  (current page.south west) --
  (current page.north west) --
  (current page.north east)}
]

\begin{tikzpicture}[rotate=90]


\draw[line width=2] (-\size,\paddedsize) -- (\size,\paddedsize);
\draw[line width=2] (-\size,-\paddedsize) -- (\size,-\paddedsize);
\draw[line width=2] (-\size,-\paddedsize) arc (270:90:\size cm and \paddedsize 
cm);
\draw[line width=2] (\size,-\paddedsize) arc (-90:90:\size cm and \paddedsize 
cm);

\draw (-\size,0) arc (180:360:\size cm and 0.5cm);
\draw (-\size,0) arc (180:0:\size cm and 0.5cm);

\draw (0,0) -- (\sigmax, \sigmay)
node[midway,right,rotate=15] {\large$r$};;

\draw (0,0) circle (\size cm);

\node[draw=none] at (+2.7,-1.3) {\Large$\Omega_{d^2}$};
\node[draw=none] at (2.7, 3.3) {\Large (a)};

\node
[label={[xshift=-0.0cm, 
yshift=-0.0cm,rotate=0]\large$\ket{\phi_{AB}}$},draw,fill=black,circle,inner
 sep=0pt,minimum size=5pt] at (-4,0) {};
 
\node[label={[xshift=0.0cm, 
yshift=-0.0cm,rotate=0]\large$\ket{\phi_A,\phi_B}$},draw,fill=black,circle,inner
sep=0pt,minimum size=5pt] at (4,0) {};

\node [label=above:$\1/d^2$,draw,fill=black,circle,inner 
sep=0pt,minimum size=5pt] at (0,0) {};

\node [label={[xshift=0.5cm, 
yshift=-0.32cm]\large$\sigma_{AB}$},fill=black,circle,inner 
sep=0pt,minimum size=5pt] at (\sigmax, \sigmay) {};

\begin{scope}[shift={(-8,5)}]
\draw[line width=2] (-\size,\paddedsize) -- (\size,\paddedsize);
\draw[line width=2] (-\size,-\paddedsize) -- (\size,-\paddedsize);
\draw[line width=2] (-\size,-\paddedsize) arc (270:90:\size cm and \paddedsize 
cm);
\draw[line width=2] (\size,-\paddedsize) arc (-90:90:\size cm and \paddedsize 
cm);

\draw (-\size,0) arc (180:360:\size cm and 0.5cm);
\draw (-\size,0) arc (180:0:\size cm and 0.5cm);

\draw (0,0) -- (\sigmax, \sigmay)
node[midway,right,rotate=15] {\large$r$};;

\draw (0,0) circle (\size cm);

\node[draw=none] at (-2.7,-1.3) {\Large$\Omega_{d}$};
 
\node[label={[xshift=0.0cm, 
yshift=-0.8cm,rotate=0]\large$\ket{\phi_{A}}$},draw,fill=black,circle,inner
sep=0pt,minimum size=5pt] at (4,0) {};

\node [label=above:$\1/d$,draw,fill=black,circle,inner 
sep=0pt,minimum size=5pt] at (0,0) {};

\node [label={[xshift=0.4cm, 
yshift=-0.4cm]\large$\rho_A$},fill=black,circle,inner 
sep=0pt,minimum size=5pt] at (\sigmax, \sigmay) {};

\node[draw=none] at (-2.7, -3.) {\Large (b)};
\end{scope}

\begin{scope}[shift={(-8,-5)}]
\draw[line width=2] (-\size,\paddedsize) -- (\size,\paddedsize);
\draw[line width=2] (-\size,-\paddedsize) -- (\size,-\paddedsize);
\draw[line width=2] (-\size,-\paddedsize) arc (270:90:\size cm and \paddedsize 
cm);
\draw[line width=2] (\size,-\paddedsize) arc (-90:90:\size cm and \paddedsize 
cm);

\draw (-\size,0) arc (180:360:\size cm and 0.5cm);
\draw (-\size,0) arc (180:0:\size cm and 0.5cm);

\draw (0,0) -- (\sigmax, \sigmay)
node[midway,right,rotate=15] {\large$r$};

\draw (0,0) circle (\size cm);

\node[draw=none] at (-2.7,-1.3) {\Large$\Omega_{d}$};
 
\node[label={[xshift=0.0cm, 
yshift=-0.8cm,rotate=0]\large$\ket{\phi_{B}}$},draw,fill=black,circle,inner
sep=0pt,minimum size=5pt] at (4,0) {};

\node [label=above:$\1/d$,draw,fill=black,circle,inner 
sep=0pt,minimum size=5pt] at (0,0) {};

\node [label={[xshift=0.4cm, 
yshift=-0.4cm]\large$\rho_B$},fill=black,circle,inner 
sep=0pt,minimum size=5pt] at (\sigmax, \sigmay) {};
\node[draw=none] at (-2.7, 3.) {\Large (c)};
\end{scope}


\draw [dash dot,-latex] (4,0) -- (4,3) arc (-360:-270:2cm) -- (-3.8,5);
\draw [dash dot,-latex] (4,0) -- (4,-3) arc (0:-90:2cm) -> (-3.8,-5);

\draw [dash dot,-latex] (-4, 0) -- (\sigmax-8+0.1,\sigmay+5-0.1);
\draw [dash dot,-latex] (-4, 0) -- (\sigmax-8+0.1,\sigmay-5+0.1);

\draw [dash dot,-latex] (\sigmax, \sigmay) -- (-8+0.1,5-0.1);
\draw [dash dot,-latex] (\sigmax, \sigmay) -- (-8+0.1,-5+0.1);

\draw [dash dot,-latex,line width=2] (0, 2.5) -- (-4, 4) 
node[midway,above,rotate=65] 
{\large$\Tr_B$};
\draw [dash dot,-latex,line width=2] (0, -2.5) -- (-4, -4) 
node[midway,above,rotate=-65] 
{\large$\Tr_A$};
\end{tikzpicture} \caption{Set of all bipartite quantum states of
dimension $d^2$, $\Omega_{d^2}$, (a) and its partial traces (b) and (c)
containing states of dimension $d$. A generic bipartite state $\sigma_{AB}$,
distant $r=3\sqrt{3}/4\pi$ from the maximally mixed state $\1/d^2$, is mapped
into $\sigma_A \approx \sigma_B \approx \1/d$, while a typical pure state
$\ket{\phi_{AB}}$ is sent into a generic mixed state $\rho_A \equiv \rho_B$
distant $r$ from $\1/d$.}\label{fig:set-of-states}
\end{figure}

To arrive at this result we considered an ensemble of normalized random density
matrices, acting on a  bipartite Hilbert space ${\mathcal H}_A \otimes
{\mathcal H}_B$, and  distributed according to the flat (Hilbert-Schmidt)
measure. Such matrices, can be generated with help of a complex Ginibre matrix
$G$ as $\rho=GG^{*}/{\rm Tr} GG^{*}$. In the simplest case of square matrices
$G$ of order $d=d_1^2$ the average trace distance of a random state $\rho$ from
the maximally mixed state $\rho_*={I }/d$ behaves asymptotically as $ ||\rho -
\rho_*||_1 \to  3\sqrt{3}/4\pi$  \cite{ppz}. However, analyzing both reduced
matrices $\rho_A={\rm Tr}_B \rho$ and $\rho_B={\rm Tr}_A \rho$ we can show that
they become $\epsilon$ close to the maximally mixed state in sense of the
operator norm, so that their smallest and largest eigenvalues do coincide. This
is visualized in Fig.~\ref{fig:set-of-states}.

 This observation implies that the state $\rho$ can be directly interpreted as 
 a Jamio{\l}kowski state $J$ representing a stochastic map $\Phi$, 
 as its partial trace $\rho_A$ is proportional to identity. Furthermore,
 as it becomes asymptotically equal to the other partial trace $\rho_B$,
 it follows that a generic  quantum channel (stochastic map) becomes 
 unital and thus bistochastic.
 
The partial trace of a random bipartite state is shown to be close to identity
provided the support of the limiting measure characterizing the bipartite state
is bounded. In particular, this holds for a family of subtract Mar\u{c}enko--Pastur 
distributions defined in Eq.~\eqref{eq:def-SMP-xy} as a free additive convolution
of two rescaled Mar\u{c}enko--Pastur distributions with different parameters 
and determining the density of a difference of two random density matrices.
In this way we could establish the upper bound for the average 
diamond norm between two channels and show
that it asymptotically converges to the lower bound $\Delta(x,y)$ given in Theorem \ref{thm:lower}.
The results obtained can be understood as an application
of the measure concentration paradigm  \cite{asz} to the space of quantum 
channels.   

\noindent \emph{Acknowledgments.} I.N.~would like to thank Anna Jen\u{c}ov\'a
and David Reeb for very insightful discussion regarding the diamond norm, which
led to several improvements and simplifications of the proof of Proposition
\ref{prop:bound-diamond}. I.N.'s research has been supported by the ANR
projects {RMTQIT}  {ANR-12-IS01-0001-01} and {StoQ} {ANR-14-CE25-0003-01}, as
well as by a von Humboldt fellowship. Financial support by the Polish National 
Science Centre under projects number 2016/22/E/ST6/00062 (ZP), 
2015/17/B/ST6/01872 ({\L}P) and 2011/02/A/ST1/00119 (K{\.Z}) is gratefully 
acknowledged.

\appendix

\section{On the partial traces of unitarily invariant random matrices}
\label{app:partial-trace-unitarily-invariant}

In this section we show a general result about unitarily invariant random matrices: under some technical convergence assumptions, the partial trace of a unitarily invariant random matrix is ``flat'', i.e.~it is close in norm to its average. 

Recall that the normalized trace functional can be extended to arbitrary permutations as follows: for a matrix $X \in M_d(\mathbb C)$, write
$$\operatorname{tr}_\pi(X):= \prod_{c \in \pi} \frac 1 d \operatorname{Tr}(X^{|c|}).$$

Recall the following definition from \cite[Section 4.3]{hpe}.
\begin{definition}
A sequence of random matrices $X_d \in M_d(\mathbb C)$ is said to have \emph{almost surely limit distribution} $\mu$ if
$$\forall p \geq 1, \quad \text{a.s.}-\lim_{d \to \infty} \operatorname{tr}(X^p) = \int x^p d\mu(x),$$
\end{definition}

\begin{proposition}\label{prop:partial-trace-flat}
Consider a sequence of hermitian random matrices $A_d \in M_{d_1(d)}(\mathbb C) \otimes M_{d_2(d)}(\mathbb C)$
and assume that
\begin{enumerate}
\item Both functions $d_{1,2}(d)$ grow to infinity, in such a way that $d_1/d_2^2 \to 0$.
\item The matrices $A_d$ are unitarily invariant.
\item The family $(A_d)$ has almost surely limit distribution $\mu$, for some compactly supported probability measure $\mu$.
\end{enumerate}
Then, the normalized partial traces $B_d:= d_2^{-1} [\operatorname{id} \otimes \operatorname{Tr}](A_d)$ converge almost surely to  multiple of the identity matrix: 
$$\text{a.s.}-\lim_{d \to \infty} \|B_d - a I_{d_1(d)}\| = 0,$$
where $a$ is the average of $\mu$:
$$a:= \int x d\mu(x).$$
\end{proposition}
\begin{theproof}
In the proof, we shall drop the parameter $d \to \infty$, but the reader should remember that the matrix dimensions $d_{1,2}$ are functions of $d$ and that all the matrices appearing are indexed by $d$. To conclude, it is enough to show that
$$\mathbb P-\lim_{d_1 \to \infty}\lambda_{\max}(B_d) = a,$$
since the statement for the smallest eigenvalue follows in a similar manner. Let us denote by 
\begin{align}
b &:= \frac{1}{d_1}\sum_{i=1}^{d_1} \lambda_i(B) = \operatorname{tr}_{(1)}(B)\\
v &:= \frac{1}{d_1}\sum_{i=1}^{d_1} (\lambda_i(B)-b)^2 = 
\frac{1}{d_1}\sum_{i=1}^{d_1} \lambda_i(B)^2 - \left[ 
\frac{1}{d_1}\sum_{i=1}^{d_1} \lambda_i(B)\right]^2 = 
\operatorname{tr}_{(12)}(B) - [\operatorname{tr}_{(1)}(B)]^2 \label{eqn:def-var}
\end{align} 
the average eigenvalue and, respectively, the variance of the eigenvalues of $B$; these are real random variables (actually, sequences of random variables indexed by $d$). 
By Chebyshev's inequality, we have a bound
\begin{equation}
\lambda_{\max}(B) \leq b + \sqrt{v} \sqrt{d_1}.
\end{equation}
Note that one could replace the $\sqrt{d_1}$ factor in the inequality above by 
$\sqrt{d_1 -1}$ by using Samuelson's inequality \cite{sam,wolkowicz1980bounds}, 
but the weaker version is enough for us. 

We shall prove now that $b \to a$ almost surely and later that $d_1 v \to 0$ almost surely, which is what we need to conclude. To do so, we shall use the Weingarten formula \cite{wei78,csn}. In the graphical formalism for the Weingarten calculus introduced in \cite{cne10a}, the expectation value of an expression involving a random Haar unitary matrix can be computed as a sum over diagrams indexed by permutation matrices; we refer the reader to \cite{cne10a} or \cite{cne16} for the details. 

Using the unitary invariance of $A$, we write $A = U \operatorname{diag}(\lambda) U^*$, for a Haar-distributed random unitary matrix $U \in \mathcal U (d_1d_2)$, and some (random) eigenvalue vector $\lambda$. Note that traces of powers of $A$ depend only on $\lambda$, so we shall write $\operatorname{tr}_\pi(\lambda):=\operatorname{tr}_\pi(A)$. We apply the Weingarten formula to a general moment of $B$, given by a permutation $\pi$:
$$\mathbb E_U \operatorname{tr}_\pi(B)=\prod_{i=1}^{\#\pi} \frac{1}{d_1} \mathbb E_U \operatorname{Tr} B^{|c_i|},$$
where $c_1, \ldots, c_{\#\pi}$ are the cycles of $\pi \in \mathcal S_p$, and $\mathbb E_U$ denotes the conditional expectation with respect to the Haar random unitary matrix $U$. From the graphical representation of the Weingarten formula \cite[Theorem 4.1]{cne10a}, we can compute the conditional expectation over $U$ (note that below, the vector of eigenvalues $\lambda$ is still random):
\begin{equation}\label{eq:Wg-B}
\mathbb E_U \operatorname{tr}_\pi (B) = d_1^{-\#\pi}d_2^{-p}\sum_{\alpha, \beta \in \mathcal S_p} d_1^{\#(\pi^{-1}\alpha)} d_2^{\#\alpha} (d_1d_2)^{\#\beta} \operatorname{tr}_\beta(\lambda) \operatorname{Wg}_{d_1d_2}(\alpha^{-1}\beta).
\end{equation}
Above, $\operatorname{Wg}$ is the Weingarten function \cite{csn} and $\operatorname{tr}_\beta(\lambda)$ is the moment of the diagonal matrix $\operatorname{diag}(\lambda)$ corresponding to the permutation $\beta$. The combinatorial factors $d_1^{\#(\pi^{-1}\alpha)}$ and $d_2^{\#\alpha}$ come from the initial wirings of the boxes respective to the vector spaces of dimensions $d_1$ (initial wiring given by $\pi$) and $d_2$ (initial wiring given by the identity permutation), see Figure \ref{fig:graphical-Wg}. The pre-factors $d_1^{-\#\pi}d_2^{-p}$ contain the normalization from the (partial) traces. Finally, the (random) factors $\operatorname{tr}_\beta(\lambda)$ are the normalized power sums of $\lambda$:
$$\operatorname{tr}_\beta(\lambda) =\prod_{i=1}^{\#\beta} (d_1d_2)^{-1}\sum_{j=1}^{d_1d_2} \lambda_j^{|w_i|},$$
where $w_1, \ldots, w_{\#\beta}$ are the cycles of $\beta$. Recall that we have assumed almost sure convergence for the sequence $(A_d)$ (and, thus, for $(\lambda_d)$):
\begin{equation}\label{eq:def-m-mu}
\forall \pi, \quad \text{a.s.}-\lim_{d \to \infty} \operatorname{tr}_\beta(\lambda) = \prod_{i=1}^{\#\beta} \int x^{|w_i|} d\mu(x) =: m_\pi(\mu).
\end{equation}
\begin{figure}[htbp]
  \centering
    \includegraphics[width=0.5\textwidth]{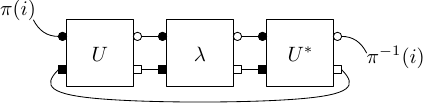}
\caption{The $i$-th group in the diagram corresponding to $m_\pi(B)$.}
 \label{fig:graphical-Wg}
\end{figure}

As a first application of the Weingarten formula \eqref{eq:Wg-B}, let us find the distribution of the random variable $b = \operatorname{Tr}(B)/d_1$. Obviously, 
\begin{equation}\label{eq:E-b}
\mathbb E_U b =\mathbb E_U \operatorname{tr}_{(1)}(B)= d_1^{-1} d_2^{-1} d_1^2d_2^2\operatorname{tr}_{(1)}(\lambda)\frac{1}{d_1d_2} = \operatorname{tr}_{(1)}(\lambda).
\end{equation}
Actually, $b$ does not depend on the random unitary matrix $U$, since 
$$b = \frac{1}{d_1} \operatorname{Tr}(B) = \frac{1}{d_1 d_2} \operatorname{Tr}(A) = \frac{1}{d_1d_2} \operatorname{Tr}(U \operatorname{diag}(\lambda)U^*) = \frac{1}{d_1d_2} \sum_{i=1}^{d_1d_2} \lambda_i = \operatorname{tr}_{(1)}(\lambda).$$
From the hypothesis \eqref{eq:def-m-mu} (with $\pi = (1)$), we have that, almost surely as $d \to \infty$, the random variable $b$ converges to the scalar $a = m_{(1)}(\mu)$. 

Let us now move on to the variance $v$ of the eigenvalues. First, we compute its expectation $\mathbb E_U v = \operatorname{tr}_{(12)}(B) - \operatorname{tr}_{(1)(2)}(B)$. We apply now the Weingarten formula \eqref{eq:Wg-B} for $\mathbb E_U\operatorname{tr}_{(12)}(B)$; the sum has $2!^2 = 4$ terms, which we compute below:
\begin{itemize}
\item $\alpha = \beta = (1)(2)$: $T_1 = d_1^2d_2^2 \operatorname{tr}_{(1)(2)}(\lambda)\frac{1}{d_1^2d_2^2-1}$
\item $\alpha = (1)(2)$, $\beta = (12)$: $T_2 = - \operatorname{tr}_{(12)}(\lambda)\frac{1}{d_1^2d_2^2-1}$
\item $\alpha = (12)$, $\beta = (1)(2)$: $T_3 = - d_1^2 \operatorname{tr}_{(12)}(\lambda)\frac{1}{d_1^2d_2^2-1}$
\item $\alpha = \beta = (12)$: $T_4 = d_1^2 \operatorname{tr}_{(12)}(\lambda)\frac{1}{d_1^2d_2^2-1}$.
\end{itemize}

Combining the expressions above with \eqref{eq:E-b}, we get
$$\mathbb E_U v = \frac{d_1^2-1}{d_1^2d_2^2-1}(\operatorname{tr}_{(12)}(\lambda) - \operatorname{tr}_{(1)(2)}(\lambda)).$$
Using the hypothesis \eqref{eq:def-m-mu}, we have thus, as $d_{1,2} \to \infty$,
$$\mathbb E v = (1+o(1)) d_2^{-2}(m_{(12)}(\mu) - m_{(1)(2)}(\mu)).$$

Let us now proceed and estimate the variance of $v$; more precisely, let us
compute $\mathbb E(v^2)$. As before, we shall compute the expectation in two
steps: first with respect to the random Haar unitary matrix $U$, and then,
using our assumption \eqref{eq:def-m-mu}, with respect to $\lambda$, in the
asymptotic limit. To perform the unitary integration, note that the Weingarten
sum is indexed by a couple $(\alpha, \beta) \in \mathcal S_4^2$, so it contains
$4!^2 = 576$ terms, see \cite{num}. In Appendix~\ref{sec:variance} we have
computed the variance of $v$ with the usage of symmetry arguments. The result,
to the first order reads
$$\operatorname{Var}(v) = \mathbb E_U(v^2) - (\mathbb E_U v)^2 = 
(1+o(1))2d_1^{-2}d_2^{-4}[m_{(12)}(\lambda) - m_{(1)(2)}(\lambda)]^2.$$
Taking the expectation over $\lambda$ and the limit (we are allowed to, by dominated convergence), we get
 $$\operatorname{Var}(v) = (1+o(1))2d_1^{-2}d_2^{-4}[m_{(12)}(\mu) - m_{(1)(2)}(\mu)]^2.$$
 
 We put now all the ingredients together:
 $$\mathbb P(\sqrt{d_1} \sqrt{v} \geq \varepsilon) = \mathbb P(v \geq \varepsilon^2 d_1^{-1}) \leq \frac{\operatorname{Var}(v)}{[\varepsilon^2d_1^{-1} - \mathbb Ev]^2} \sim \frac{Cd_1^{-2}d_2^{-4}}{[\varepsilon^2 d_1^{-1} - (1+o(1))C'd_2^{-2}]^2},$$
 where $C,C'$ non-negative constants depending on the limiting measure $\mu$. Using $d_1 \ll d_2^2$, the dominating term in the denominator above is $\varepsilon^2 d_1^{-1}$, and thus we have:
 $$\mathbb P(\sqrt{d_1} \sqrt{v} \geq \varepsilon) \lesssim C\varepsilon^{-4}d_2^{-4}.$$
 Since the series $\sum d_2^{-4}$ is summable, we obtain the announced almost sure convergence by the Borel-Cantelli lemma, finishing the proof.
\end{theproof}

\section[Calculation of the variance Var(v)]{Calculation of the variance $\operatorname{Var}(v)$ 
}\label{sec:variance}
In this appendix we compute the centered second moment of the variable $v$ 
defined in~\eqref{eqn:def-var} necessary to show almost sure convergence $d_1 v 
\to 0$.
We remind here, that $A \in M_{d_1}(\C) \otimes M_{d_2}(\C)$ and 
$B = d_2^{-1} [\operatorname{id}\otimes \tr ] (A)$.
Because we assume that $A$ has unitarly invariant distribution, we can write 
\begin{equation}
A = U \diag(\lambda) U^\dagger = \sum_{i=0}^{d_1 d_2 - 1} 
\lambda_i \proj{U_i},
\end{equation}
where $\ket{U_i} = U \ket{i}$ is $i$-th column of matrix $U$ and 
\begin{equation}
B = d_2^{-1} \tr_2 A =  d_2^{-1} \sum_{i=0}^{d_1 d_2 - 1} \lambda_i \tr_2 
\proj{U_i}.
\end{equation}
We denote $\rho_i = \tr_2 \proj{U_i}$ and consider mixed moments computed in 
Lemma~\ref{lemma:mixed-miments}
\begin{equation} \label{eqn:moments-definition}
\mathcal{M}(i,j,k,l) = \E_U \tr (\rho_i \rho_j) \tr (\rho_k \rho_l),
\end{equation}
where $\E_U$ denotes the  conditional expectation with respect to the Haar 
random unitary matrix $U$. We also define symmetric mixed moments 
\begin{equation}\label{eqn:sym-moments-definition}
\mathcal{SM}(i,j,k,l) = \E_U \tr (\rho_i \rho_j) \tr (\rho_k \rho_l)
- \E_U \tr (\rho_i \rho_j)  \ \E_U \tr (\rho_k \rho_l).
\end{equation}

\begin{proposition}\label{prop:var-calc}
Let $v =  \frac{1}{d_1}\tr B^2 - \left(\frac{1}{d_1}\tr B \right)^2$. Denoting 
$\operatorname{Var}_U (v) = \E_U v^2 - (\E_U v)^2$, we have
\begin{equation}
\operatorname{Var}_U (v) = \frac{2 (\mu_1^2 - \mu_2)^2}{d_1^2 d_2^4}(1+ o(1))
\end{equation}
as $d_1,d_2 \to \infty$, in the above $\mu_k = \frac{1}{d_1 d_2} \sum 
\lambda_i^k$.
\end{proposition}
Direct computations with the usage of symmetric moments $\mathcal{SM}$ give us
\begin{equation}\label{eqn:var-calculation}
\begin{split}
\operatorname{Var}_U (v) &=
\frac{1}{d_1^2 d_2^4} \Big[
(d_1 d_2)^4 \mu_1^4 \mathcal{SM}(0,1,2,3) \\
& +
2 (d_1 d_2)^3 \mu_2 \mu_1^2 \Big(\mathcal{SM}(0,0,1,2) + 2 
\mathcal{SM}(0,1,0,2) -3 \mathcal{SM}(0,1,2,3) \Big) \\
& +
4 (d_1 d_2)^2 \mu_3 \mu_1 \Big( \mathcal{SM}(0,0,0,1) - \mathcal{SM}(0,0,1,2)
-2 \mathcal{SM}(0,1,0,2) + 2 \mathcal{SM}(0,1,2,3) \Big) \\
& +
(d_1 d_2)^2 \mu_2^2 \Big( \mathcal{SM}(0,0,1,1) - 2 \mathcal{SM}(0,0,1,2) 
+ 2 \mathcal{SM}(0,1,0,1)-4 \mathcal{SM}(0,1,0,2)+3\mathcal{SM}(0,1,2,3) \Big) 
\\
& +
d_1 d_2 \mu_4 \Big(\mathcal{SM}(0,0,0,0) -4 \mathcal{SM}(0,0,0,1) 
-\mathcal{SM}(0,0,1,1) + 4 \mathcal{SM}(0,0,1,2) -2\mathcal{SM}(0,1,0,1) \\
&  \ \ \  \ \  \ \  \ \  \ \ \ \  \ + 8 \mathcal{SM}(0,1,0,2)-6 
\mathcal{SM}(0,1,2,3) \Big)
\Big] \\
&=
\frac{2 \left(d_1^2-1\right) \left(d_2^2-1\right) }{d_2^2 \left(d_1^2 
   d_2^2-1\right){}^2 \left(d_1^4 d_2^4-13 d_1^2 d_2^2+36\right)}
\Big(d_1^4 d_2^4 \left(\mu _1^2-\mu _2\right){}^2 
\\
&
 \ \ \ \ \ \ \ +d_1^2 d_2^2 \left(11 \mu 
_1^4-22 \mu _2 \mu _1^2+20 \mu _3 \mu _1-4 \mu _2^2-5 \mu _4\right)+5
   \left(3 \mu _2^2-4 \mu _1 \mu _3+\mu _4\right)\Big).
\end{split}
\end{equation}
The above formula gives the exact result for $\operatorname{Var}_U(v)$. 
Considering 
the limiting behaviour $d_1,d_2 \to \infty$ we get
\begin{equation}
\operatorname{Var}_U (v) = \frac{2 (\mu_1^2 - \mu_2)^2}{d_1^2 d_2^4}(1+ o(1)),
\end{equation}
which completes the proof of Proposition~\ref{prop:var-calc}.

The moments computed here are used in equation~\eqref{eqn:var-calculation} to 
obtain the variance $\operatorname{Var}_U(v)$.
\begin{lemma} \label{lemma:mixed-miments}
We have the following formulas for mixed moments defined in 
equation~\eqref{eqn:moments-definition}. Note, that because of symmetry we 
cover all possible cases.
\begin{equation}
\begin{split}
\mathcal{M}(0,0,0,0) &= \frac{d_2 d_1^3+2 \left(d_2^2+2\right) d_1^2+d_2
					     \left(d_2^2+10\right) d_1+4 d_2^2+2}{\left(d_1 
					     d_2+1\right) \left(d_1 d_2+2\right) \left(d_1 
					     d_2+3\right)},    \\
\mathcal{M}(0,0,0,1)&= \frac{\left(d_2^2-1\right) \left(d_1 
						\left(d_1+d_2\right) \left(d_1 d_2+ 
						4\right)+2\right)}{\left(d_1 d_2-1\right) \left(d_1  
						d_2+1\right) \left(d_1 d_2+2\right) \left(d_1 
						d_2+3\right)},\\
\mathcal{M}(0,0,1,1) &= \frac{\left(d_1 d_2 \left(d_1 d_2+2\right)-4\right)
						 \left(d_1+d_2\right){}^2+4}{d_1 d_2 \left(d_1 
						 d_2-1\right)\left(d_1 d_2+2\right) \left(d_1 
						 d_2+3\right)},      \\
\mathcal{M}(0,0,1,2) &= \frac{\left(d_2^2-1\right) \left(d_1 
						 \left(d_1+d_2\right) \left(d_1 d_2 \left(d_1 
						 d_2+4\right)+2\right)-2\right)}{d_1 d_2 \left(d_1    
						 d_2-1\right) \left(d_1 d_2+1\right) \left(d_1 
						 d_2+2\right)\left(d_1 d_2+3\right)},       \\
\mathcal{M}(0,1,0,1) &= \frac{\left(d_2^2-1\right) \left(d_1 \left(6 
d_2+d_1 	
						 \left(d_2^2\left(d_1 d_2+5\right) -2\right) 
						 \right)+2\right)}{d_1 d_2 \left(d_1 d_2-1\right) 
						 \left(d_1 d_2+1\right) \left(d_1 d_2+2\right) 
						 \left(d_1 d_2+3\right)}, \\
\mathcal{M}(0,1,0,2) &= \frac{\left(d_2^2-1\right) \left(d_1 \left(d_1 
						 \left(d_2 \left(d_1  \left(3 d_2^2+d_1 
						 \left(d_2^2-1\right) d_2-4\right)-3 d_2\right) +2 
						 \right)-8 d_2\right)-2\right)}{d_1 d_2 \left(d_1   
						 d_2-2\right) \left(d_1 d_2-1\right) \left(d_1 d_2+1 
						 \right) \left(d_1 d_2+2\right) \left(d_1 d_2+3\right)} 
						 \\
\mathcal{M}(0,1,2,3) &=  \frac{\left(d_2^2-1\right) \left(d_2^2 		
						  \left(d_2^2-1\right) d_1^4+2 \left(7-6 d_2^2\right) 
						  d_1^2+22\right)}{d_1^2 d_2^2 \left(d_1^2 d_2^2 - 7 
						  \right){}^2-36}.
\end{split}
\end{equation}
\end{lemma}
The rest of this section is devoted to the proof of the above lemma. 
We will omit the subscript $U$ in the expectation, because matrices $\rho_i$
depend only on the Haar unitary matrix $U$. Following the result of 
Giraud~\cite{giraud2007distribution} we find the second moment of the purity 
\begin{equation}
\mathcal{M}(0,0,0,0)=\E (\tr \rho_0^2)^2 
= \frac{d_2 d_1^3+2 \left(d_2^2+2\right) d_1^2+d_2
   \left(d_2^2+10\right) d_1+4 d_2^2+2}{\left(d_1 d_2+1\right)
   \left(d_1 d_2+2\right) \left(d_1 d_2+3\right)},
\end{equation}
where $\rho_0 = \tr_2 \ketbra{U_0}{U_0}$.

Next we consider the moments $\mathcal{M}(0,0,0,1)$ defined 
in~\eqref{eqn:moments-definition}
\begin{equation}
\begin{split}
\mathcal{M}(0,0,0,1) &=\E \tr \rho_0^2 \tr(\rho_0\rho_1) 
=
\frac{1}{d_1 d_2 - 1} \E \tr\rho_0^2 \tr(\rho_0 (d_2 \1_{d_1} -\rho_0)).
\end{split}
\end{equation}
The above follows from the fact, that we have invariance with respect to the 
permutation of columns of $U$, and therefore $\E \tr \rho_0^2 \tr(\rho_0\rho_1) 
= \E \tr \rho_0^2 \tr(\rho_0\rho_2)$. Next we note, that $\sum_{i=0}^{d_1 
d_2-1} \rho_i = d_2 \1_{d_1}$. Using the above we obtain  
\begin{equation}
\begin{split}
\mathcal{M}(0,0,0,1) &=
\frac{1}{d_1 d_2 - 1} \left(d_2 \E \tr \rho_0^2  - \E (\tr \rho_0^2 )^2 
\right) \\
&=
\frac{1}{d_1 d_2 - 1} \left(d_2 \frac{d_1+d_2}{d_1 d_2+1}  - 
\mathcal{M}(0,0,0,0) \right) 
\\
&= \frac{\left(d_2^2-1\right) \left(d_1 \left(d_1+d_2\right) \left(d_1
   d_2+4\right)+2\right)}{\left(d_1 d_2-1\right) \left(d_1
   d_2+1\right) \left(d_1 d_2+2\right) \left(d_1 d_2+3\right)}.
\end{split}
\end{equation}

In order to get other mixed moments we need to perform another integration.
We start with expectations of the following kind 
\begin{equation}
\mathcal{M}(0,0,1,1) = \E \tr \rho_0^2 \tr \rho_1^2 = 
\E \tr (\tr_2 \proj{U_0})^2 \tr (\tr_2 \proj{U_1})^2.
\end{equation}
Note that if we multiply matrix $U$ by a unitary  matrix which does not change 
the first column we will not change the expectation value. In fact we can 
integrate over the subgroup of matrices which does not change the first column 
of $U$. Now for a moment we fix the matrix $U$ and consider the expectation 
value
\begin{equation}
\tr (\tr_2 \proj{U_0})^2  \E_{V} \tr (\tr_2 U V \proj{1} V^{\dagger} U^{\dagger}
)^2,
\end{equation}
where matrices $V$ are in the form 
\begin{equation}
V = \left(
\begin{smallmatrix}
1 & 0 & 0 & \dots & 0 \\
0 & v_{1,1} & v_{1,2} & \dots &v_{1,d_1 d_2-1} \\
0 & v_{2,1} & v_{1,2} & \dots &v_{2,d_1 d_2-1} \\
\vdots & \vdots & \vdots & \ddots &\vdots\\
0 & v_{d_1 d_2-1,1} & v_{d_1 d_2-1,2} & \dots &v_{d_1 d_2-1,d_1 d_2-1} \\
\end{smallmatrix}
\right).
\end{equation}
The $\E_V$ is an expectation with respect to the Haar measure on $U(d_1 d_2-1)$ 
embedded in $U(d_1 d_2)$, in the above way. Note, that the vector $U V \ket{1}$
represents a random orthogonal vector to the $\ket{U_0} = U \ket{0}$.

First we calculate 
\begin{equation}
\begin{split}
&
\E_{V} (U V \proj{1} V^{\dagger} U^{\dagger})\otimes (U V \proj{1} V^{\dagger} 
U^{\dagger}) \\
&=\E_{V} (U \proj{V_1}U^{\dagger})\otimes (U \proj{V_1} U^{\dagger}) \\
&= (U \otimes U) \E_{V} \proj{V_1}\otimes \proj{V_1} (U \otimes U)^\dagger.
\end{split}
\end{equation}
Now, using standard integrals we obtain
\begin{equation}
\E_{V} \proj{V_1}\otimes \proj{V_1} =
\frac{1}{(d_1 d_2-1)d_1 d_2}\sum_{i_1 i_2 j_1 j_2}
\left(\delta_{i_1 j_1} \delta_{i_1 j_1} + \delta_{i_1 j_2} 
\delta_{i_2 j_1} \right) \theta(i_1 i_2 j_1 j_2) \ketbra{i_1 i_2}{j_1 j_2}.
\end{equation}
where $\theta(x) = (1 - \delta_{x,0})$ and incorporates the condition that 
first 
element of vector $\ket{V_1}$ is zero.
Now we obtain, after elementary calculations, using the fact that $U$ is 
unitary
\begin{equation}
\begin{split}
&(U \otimes U) \E_{V} \proj{V_1}\otimes \proj{V_1} (U \otimes U)^\dagger \\
&=
\frac{1}{(d_1 d_2-1)d_1 d_2}
\sum_{i_1 i_2 j_1 j_2}
\Big(
(\delta_{i_1 j_1} - u_{i_1,0} u_{j_1,0})
(\delta_{i_2 j_2} - u_{i_2,0} u_{j_2,0}) 
\\
&  \ \ \ \ \ \ \ \ \ \ \ \ \ \ \ \ \ 
+
(\delta_{i_1 j_2} - u_{i_1,0} u_{j_2,0})
(\delta_{i_2 j_1} - u_{i_2,0} u_{j_1,0})
\Big) \theta(i_1 i_2 j_1 j_2) \ketbra{i_1 i_2}{j_1 j_2} \\
&=
\frac{1}{(d_1 d_2-1)d_1 d_2}
\Big(
\1_{d_1 d_2 d_1 d_2} + \proj{U_0} \otimes \proj{U_0} - \1_{d_1 d_2} \otimes 
\proj{U_0} - 
\proj{U_0} \otimes \1_{d_1 d_2} \\
& \ \ \ \ \ \ \  + S_{d_1 d_2} (\1_{d_1 d_2d_1 d_2} + \proj{U_0} 
\otimes 
\proj{U_0} - 
\1_{d_1 d_2} \otimes 
\proj{U_0} - 
\proj{U_0} \otimes \1_{d_1 d_2})
\Big),
\end{split}
\end{equation}
where $S_{N}$ is a swap operation on two systems of dimensions 
$N$ each, i.e. $S = \sum_{i_1,i_2=0}^{N-1} \ketbra{i_1i_2}{i_2 i_1}$. So we get 
\begin{equation}
\begin{split}
&(U \otimes U) \E_{V} \proj{V_1}\otimes \proj{V_1} (U \otimes U)^\dagger \\
&=
\frac{1}{(d_1 d_2-1)d_1 d_2}
(\1_{d_1 d_2d_1 d_2} + S_{d_1 d_2})\Big(
\1_{d_1 d_2} + \proj{U_0} \otimes \proj{U_0} - \1_{d_1 d_2} \otimes \proj{U_0} 
- \proj{U_0} \otimes \1_{d_1 d_2} \Big),
\end{split}
\end{equation}
We are going to use several times the following identity often used in quantum 
information. For two square  matrices $\rho_1,\rho_2$ of size $N$
\begin{equation}
\tr \rho_1 \rho_2 = \tr S_N (\rho_1 \otimes \rho_2 ).
\end{equation}
This identity allows us to obtain  
\begin{equation}
\begin{split}
&\E_{V} \tr (\tr_2 U \proj{V_1}U^{\dagger})^2 =
\E \tr S_{d_1}(\tr_2 U \proj{V_1} U^{\dagger}) \otimes 
(\tr_2 U \proj{V_1} U^{\dagger})
\\
&=
\E_{V} \tr S_{d_1}\tr_{2,4} (U \proj{V_1} U^{\dagger} \otimes 
U \proj{V_1} U^{\dagger}) \\
&=
\tr S_{d_1}\tr_{2,4} (U \otimes U) \E_{V} \proj{V_1}\otimes \proj{V_1} (U 
\otimes U)^\dagger 
\\
&=\frac{1}{(d_1 d_2-1)d_1 d_2} \tr S_{d_1}\tr_{2,4}  
(\1_{d_1 d_2 d_1 d_2} + S_{d_1 d_2})\Big(
\1_{d_1 d_2 d_1 d_2} + \proj{U_0} \otimes \proj{U_0} \\
& \ \ \  \ \ - \1_{d_1 d_2} \otimes \proj{U_0} 
- \proj{U_0} \otimes \1_{d_1 d_2} \Big).
\end{split}
\end{equation}
After performing partial trace over subsystems 2 and 4 we get 
\begin{equation}
\begin{split}
&\E_{V} \tr (\tr_2 U \proj{V_1}U^{\dagger})^2 \\
&=\frac{1}{(d_1 d_2-1)d_1 d_2} 
\Big(
d_1 d_2^2 + \tr \rho_0^2 -  d_2 \tr \rho_0 - d_2 \tr \rho_0
+ d_1^2 d_2 + \tr (\rho'_0)^2 -  d_1 \tr \rho'_0 - d_1 \tr \rho'_0 \Big) \\
&=
\frac{1}{(d_1 d_2-1)d_1 d_2}  
\Big(
d_1 d_2^2 + d_1^2 d_2 - 2 d_1 - 2 d_2 + 2 \tr \rho_0^2 \Big).
\end{split}
\end{equation}
In the above formulas we  used $ \rho'_0 = \tr_1 \proj{U_0}$ and the fact, that
two partial traces of a pure bi-partite state have the same purity $\tr 
\rho_0^2 =  \tr (\rho'_0)^2$.
Using the above we find the desired expectation 
\begin{equation}
\begin{split}
\mathcal{M}(0,0,1,1) &= \E \tr \rho_0^2  \tr \rho_1^2 =
\E \tr \rho_0^2  \E_{V} \tr (\tr_2 U \proj{V_1}U^{\dagger})^2 \\
&=
\frac{1}{(d_1 d_2-1)d_1 d_2}  
\Big(
(d_1 d_2^2 + d_1^2 d_2 - 2 d_1 - 2 d_2 ) \E \tr \rho_0^2  + 2 \E (\tr 
\rho_0^2)^2 \Big)  \\
&=
\frac{1}{(d_1 d_2-1)d_1 d_2}  
\Big(
(d_1 d_2^2 + d_1^2 d_2 - 2 d_1 - 2 d_2 ) \frac{d_1+d_2 }{d_1 d_2 + 1} +
 2  \mathcal M(0,0,0,0) \Big)  \\
&= 
\frac{\left(d_1 d_2 \left(d_1 d_2+2\right)-4\right)
   \left(d_1+d_2\right){}^2+4}{d_1 d_2 \left(d_1 d_2-1\right)
   \left(d_1 d_2+2\right) \left(d_1 d_2+3\right)}.
\end{split}
\end{equation}

Using inner integral we can also calculate the other mixed moments  
\begin{equation}
\begin{split}
&\mathcal{M}(0,1,0,1) = \E \tr \rho_0 \rho_1  \tr \rho_0 \rho_1 =
\E \tr (\rho_0 \otimes \rho_0 )  (\rho_1 \otimes \rho_1) \\
&=
\E \tr (\rho_0 \otimes \rho_0 )
\tr_{2,4}
(U \otimes U) \E_{V} \proj{V_1}\otimes \proj{V_1} (U \otimes U)^\dagger \\
&=
\frac{1}{(d_1 d_2-1)d_1 d_2}  
\E \tr (\rho_0 \otimes \rho_0 )
\tr_{2,4}(
\1_{d_1 d_2 d_1 d_2} + S_{d_1 d_2,d_1 d_2})\\
& \ \ \  \times
\Big(
\1_{d_1 d_2 d_1 d_2} + \proj{U_0} \otimes \proj{U_0} - \1_{d_1 d_2} \otimes 
\proj{U_0} - \proj{U_0} \otimes \1_{d_1 d_2} \Big)\\
&=
\frac{1}{(d_1 d_2-1)d_1 d_2}\E 
\Big(\tr (\rho_0 \otimes \rho_0 )
(d_2^2 \1_{d_1^2} + \rho_0 \otimes \rho_0 - d_2 \1_{d_1} \otimes \rho_0 - d_2 
\rho_0 \otimes \1_{d_2} ) \\ 
&+
\tr S_{d_1 d_2,d_1 d_2} \Big(
\1_{d_1 d_2 d_1 d_2} + \proj{U_0} \otimes \proj{U_0} - \1_{d_1 d_2} \otimes 
\proj{U_0} - \proj{U_0} \otimes \1_{d_1 d_2} \Big) (\rho_0 \otimes \1_{d_2} 
\otimes \rho_0 \otimes \1_{d_{2}})
\Big)
\\
&=
\frac{1}{(d_1 d_2-1)d_1 d_2}\E 
\Big(
d_2^2 +(\tr\rho_0^2)^2 - 2 d_2\tr \rho_0 \tr\rho_0^2
\\
&+
\tr (\rho_0^2 \otimes \1_{d_2} )
+
\tr (\proj{U_0} \rho_0 \otimes \1_{d_2} )^2
- 2 \tr (\rho_0 \otimes \1_{d_2})(\proj{U_0} \rho_0 \otimes \1_{d_2})
\Big)\\
&=
\frac{1}{(d_1 d_2-1)d_1 d_2}\E 
\Big(
d_2^2 -2d_2 \tr\rho_0^2+(\tr\rho_0^2)^2
+
d_2 \tr \rho_0^2
+
(\tr\rho_0^2)^2
-2 \tr \rho_0^3
\Big)\\
&=
\frac{1}{(d_1 d_2-1)d_1 d_2}
\Big(
d_2^2 - d_2 \frac{d_1+d_2}{d_1 d_2+1}
+2 \mathcal M (0,0,0,0)
-2 \E \tr \rho_0^3
\Big)\\
&=
\frac{\left(d_2^2-1\right) \left(d_1 \left(6 d_2+d_1 \left(d_2^2
   \left(d_1 d_2+5\right)-2\right)\right)+2\right)}{d_1 d_2
   \left(d_1 d_2-1\right) \left(d_1 d_2+1\right) \left(d_1
   d_2+2\right) \left(d_1 d_2+3\right)}.
\end{split}
\end{equation}
This is because $ \E \tr \rho_0^3 =  \frac{\left(d_1+d_2\right){}^2+d_1 
d_2+1}{\left(d_1 d_2+1\right) \left(d_1 d_2+2\right)} $ 
see~\cite{sommers2004statistical}.

Using above results we obtain other moments 
\begin{equation}
\begin{split}
\mathcal{M}(0,0,1,2) &=\E \tr\rho_0^2 \tr(\rho_1 \rho_2) 
=
\frac{1}{d_1 d_2-2} \E \tr \rho_0^2 \tr(\rho_1 (d_2 \1_{d_1} -\rho_0 -\rho_1)) 
\\
&=
\frac{1}{d_1 d_2-2} \left(d_2 \E \tr \rho_0^2  -
\E \tr \rho_0^2 \tr (\rho_1 \rho_0) -
\E \tr \rho_0^2 \tr \rho_1^2 
\right) \\
&=
\frac{1}{d_1 d_2-2} \left(
d_2 \frac{d_1 + d_2}{d_1 d_2 +1} 
- \mathcal{M}(0,0,0,1)
- \mathcal{M}(0,0,1,1)
\right) \\
&=
\frac{\left(d_2^2-1\right) \left(d_1 \left(d_1+d_2\right) \left(d_1
   d_2 \left(d_1 d_2+4\right)+2\right)-2\right)}{d_1 d_2 \left(d_1
   d_2-1\right) \left(d_1 d_2+1\right) \left(d_1 d_2+2\right)
   \left(d_1 d_2+3\right)}.
\end{split}
\end{equation}

Next we consider the mixed moment of type $(0,1,0,2)$,
\begin{equation}
\begin{split}
\mathcal{M}(0,1,0,2) &=
\E \tr(\rho_0 \rho_1) \tr(\rho_0 \rho_2) 
=
\frac{1}{d_1 d_2-2} \E \tr(\rho_0\rho_1) \tr(\rho_0 (d_2 \1_{d_1} 
-\rho_0 -\rho_1)) \\
&=
\frac{1}{d_1 d_2-2} 
\left(
d_2 \E \tr(\rho_0\rho_1) 
- 
\E \tr(\rho_0\rho_1) \tr(\rho_0^2)
- \E \tr(\rho_0\rho_1) \tr(\rho_0 \rho_1)
\right)\\
&=
\frac{\left(d_2^2-1\right) \left(d_1 \left(d_1 \left(d_2 \left(d_1
   \left(3 d_2^2+d_1 \left(d_2^2-1\right) d_2-4\right)-3
   d_2\right)+2\right)-8 d_2\right)-2\right)}{d_1 d_2 \left(d_1
   d_2-2\right) \left(d_1 d_2-1\right) \left(d_1 d_2+1\right)
   \left(d_1 d_2+2\right) \left(d_1 d_2+3\right)}.
\end{split}
\end{equation}


Consider now the last case of all different indices
\begin{equation}
\begin{split}
\mathcal{M}(0,1,2,3) &=\E \tr(\rho_0 \rho_1) \tr(\rho_2 \rho_3)\\
&=
\frac{1}{d_1 d_2-3}\E \tr(\rho_0 \rho_1) \tr(\rho_2(d_2 \1_{d_1} - 
\rho_0 - \rho_1  - \rho_2)\\
&=
\frac{1}{d_1 d_2-3}\E \tr(\rho_0\rho_1) 
\left(
d_2 
- \tr \rho_2 \rho_0
- \tr \rho_2 \rho_1
- \tr \rho_2 \rho_2
\right) \\
&=
\frac{\left(d_2^2-1\right) \left(d_2^2 \left(d_2^2-1\right) d_1^4+2
   \left(7-6 d_2^2\right) d_1^2+22\right)}{d_1^2 d_2^2 \left(d_1^2
   d_2^2-7\right){}^2-36}.
\end{split}
\end{equation}

In this way we calculated all the moments defined in 
eqn.~\eqref{eqn:moments-definition}. Symmetrizing them according the 
eqn.~\eqref{eqn:sym-moments-definition} they can be used in 
eqn.~\eqref{eqn:var-calculation} to establish Proposition~\ref{prop:var-calc}.

\end{document}